\documentclass[12pt]{article}

\usepackage{graphicx}
\usepackage[top=1in,bottom=1in,left=1in,right=1in,paper=a4paper]{geometry}
\usepackage{lineno}
\usepackage{multirow}
\usepackage{chemfig}
\usepackage{chemformula}
\usepackage{amsmath,bm}
\usepackage{amssymb}
\usepackage{wasysym}
\usepackage{braket}
\usepackage{xcolor}
\usepackage{caption}
\usepackage{subcaption}
\usepackage[normalem]{ulem}

\usepackage[style=nature]{biblatex}
\title{ Radical pairs and superoxide amplification can explain magnetic field effects on planarian regeneration}
\author{{\normalsize Rishabh\textsuperscript{1,2,3,*}, Hadi Zadeh-Haghighi\textsuperscript{1,2,3}, Christoph Simon\textsuperscript{1,2,3,\dag}}\\\\
{\footnotesize\textsuperscript{1}Department of Physics and Astronomy, University of Calgary, Calgary, Alberta, Canada}\\
{\footnotesize\textsuperscript{2}Institute for Quantum Science and Technology, University of Calgary, Calgary, Alberta, Canada}\\
{\footnotesize\textsuperscript{3}Hotchkiss Brain Institute, University of Calgary, Calgary, Alberta, Canada}\\
{\footnotesize\textsuperscript{*}rishabh1@ucalgary.ca; \textsuperscript{\dag}csimo@ucalgary.ca}}
\date{{\footnotesize\today}}

\begin{document}

\maketitle

\section*{Abstract}
Weak magnetic field exposure can affect many biological processes across a wide range of living organisms. Recently, it has been observed that weak magnetic fields can modulate reactive oxygen species (ROS) concentration, affecting regeneration in planaria. These effects show unusual nonlinear dependence on magnetic field strength, including a sign change. In another study by the same group, superoxide is identified as the particular ROS being modulated. We propose a radical pair mechanism based on a flavin-superoxide radical pair to explain the modulation of superoxide production and its effect on planarian regeneration. The results of our calculations favor a triplet-born radical pair. Our yield calculations can reproduce the observed magnetic field dependence, including the sign change. Moreover, to explain the size of the effect on ROS concentration, we suggest a simple amplification model inspired by known biochemical mechanisms  and lay out the conditions for such a model to work. Further, we also make empirical predictions concerning the hypomagnetic field effects on planarian regeneration.

\section{Introduction}
Hundreds of studies exist that show exposure to weak magnetic fields (WMFs) can influence many biological processes in various organisms, from bacteria to human beings~\cite{zadeh2022magnetic}. As the corresponding energies for WMFs are far smaller than thermal energies at room temperature, no obvious classical explanation is available. In recent years, the radical pair mechanism (RPM)~\cite{closs1969mechanism,steiner1989magnetic} has been proposed as a quantum mechanical explanation for several WMF effects in biology~\cite{hore2016radical,zadeh2022magnetic}. 
\par
The RPM involves the simultaneous creation of a pair of radicals. A radical is a molecule that contains an unpaired electron. The spins of the two unpaired electrons, one on each constituent molecule of the radical pair (RP), evolve coherently. RPs usually start in either singlet or triplet initial states. These RPs interact with the nuclear spins in the vicinity of each radical via hyperfine (HF) interactions and with an external magnetic field via the Zeeman interaction. Because of HF interactions, neither singlet nor triplet states are stationary states of the spin Hamiltonian. A RP in such a non-stationary superposition evolves coherently and leads to singlet-triplet interconversion, whose frequency is determined by the HF and Zeeman interactions. Altering the Zeeman interaction (by changing the external magnetic field) or HF interaction (by substituting an isotope with a different spin) can change this singlet-triplet interconversion. A characteristic feature of the RPM is that it involves the formation of spin-selective chemical products from singlet and triplet states; therefore, any change in the singlet-triplet interconversion results in altered yields of chemical products formed via the RPM.
\par
Reactive oxygen species (ROS) are derivatives of oxygen and include free radicals and non-radical species. Superoxide (\ch{O2^{.-}}) and hydrogen peroxide (\ch{H2O2}) are two of the most biologically important members of the ROS family. ROS are vital for various cellular processes. Unusually high ROS levels are related to multiple pathological conditions, including aging, cardiovascular and respiratory dysfunction, cancer, and several neurodegenerative diseases~\cite{davalli2016ros,al2011oxidases,weinberg2019reactive,singh2019oxidative}. Low ROS levels, conversely, can cause impaired learning and memory, defective cellular proliferation and differentiation, among other detrimental effects~\cite{kishida2006synaptic,mofarrahi2008regulation,vieira2011modulation}. Studies have also shown the importance of ROS in cellular signaling~\cite{sies2020reactive,d2007ros,sinenko2021physiological}.
\par
Researchers have shown in multiple scenarios that the cellular production of ROS is magnetically sensitive~\cite{wang2017magnetic,calabro2013effects,martino2011modulation,poniedzialek2013reactive}. In many other studies involving magnetic field effects on higher-level processes such as regeneration, vasculogenesis, apoptosis, and Doxorubicin-induced toxicity, it has been shown that these effects are mediated by modulating ROS concentration, i.e., ambient magnetic field changes result in altered ROS levels, which in turn have downstream effects~\cite{zhang2021long,bekhite2010static,de2006magnetic,hajipour2018static}.
\par
RPM-based models have been proposed to explain a few experiments involving WMF effects on ROS production. Usselman et al. proposed a flavin and superoxide-based RPM to explain the effects of oscillating magnetic fields at Zeeman resonance on ROS production in human umbilical vein endothelial cells~\cite{usselman2016quantum}. In a later study by our group, a similar mechanism was used to explain the modulation of ROS production and the related attenuation of adult hippocampal regeneration in mice in a hypomagnetic field environment~\cite{rishabh2022radical}. Singlet-born RP was found to agree with the experimental results, whereas predicted yields of \ch{O2^{.-}} for triplet-born RP contradict observed values.
\par
Planaria are flatworms with a large number of somatic stem cells called neoblasts, which account for nearly a quarter of their total cell population~\cite{baguna2012planarian}. Due to this large adult stem cell population, they have an astonishing capability for regenerating all tissues, including the central nervous system~\cite{cebria2007regenerating}. 
\par
Studies on various organisms, including on planaria, have shown that regeneration following an injury is intimately connected with the process of apoptosis~\cite{pellettieri2010cell,metzstein1998genetics}. Apoptosis is a programmed cell death, an evolutionarily conserved process~\cite{guerin2021cell}. Apoptosis regulates cell number and proliferation during regeneration. After an injury, two bursts in the rate of apoptosis are observed. The first burst, which occurs early in the process and near the wound site, stimulates regeneration, and a later burst regulates regenerative patterning. The first apoptotic peak occurs 1 to 4 hours after injury, whereas the second peak occurs 3 days post-injury~\cite{beane2013bioelectric}. ROS are known to play an important role in regeneration in general and in planaria in particular. ROS plays a vital role in apoptosis induction~\cite{simon2000role}. In zebrafish, ROS production has been shown to trigger apoptosis-induced compensatory proliferation required for regeneration to proceed~\cite{gauron2013sustained}. In planaria, the amputation of the head and tail induces a ROS burst at the wound site, and inhibition of ROS production causes regeneration defaults, including reduced blastemas~\cite{van2019weak}. A blastema is a collection of unpigmented adult stem cell progeny that forms the core of new tissues.
\par
The strong dependence of planarian regeneration on ROS levels, in conjunction with the fact that WMFs in various biological contexts can alter ROS concentrations, suggests a high possibility that planarian regeneration is itself sensitive to WMFs and that these effects are mediated via changes in ROS levels. Van Huizen et al. conducted studies to this effect and found that planarian regeneration is indeed sensitive to WMFs~\cite{van2019weak}. They reported that WMFs altered stem cell proliferation and subsequent differentiation by changing ROS accumulation and downstream heat shock protein 70 expression.
\par
Although Van Huizen et al.~\cite{van2019weak} established ROS-mediated WMF effects on planarian regeneration, the specific ROS involved remained a mystery. In a later study by the same group, they identified \ch{O2^{.-}} as a specific ROS being modulated~\cite{kinsey2023weak}. They also suggested that their observations can be explained by the RPM.
\par
In this paper, we analyze Kinsey et al.'s~\cite{kinsey2023weak} suggestion that the RPM can explain their observations regarding WMF effects on ROS levels. We start with an RPM-based model similar to that used in the previous studies~\cite{usselman2016quantum,rishabh2022radical} on WMF effects on ROS and use master equation analysis~\cite{haberkorn1976density,steiner1989magnetic} to calculate the \ch{O2^{.-}} yield. We found that the theoretical predictions of our model can mimic the observed behavior as a function of the magnetic field, including the change of sign. However, the amplitude of the effects was smaller than that of the observed variation in superoxide, approximately by an order of magnitude. To explain the amplitude of effects, we suggest a simple amplification model inspired by biology and lay out the necessary conditions for such a model to work. In this study, we considered both singlet-born and triplet-born RPs as well as F-pairs, i.e., RPs resulting from random encounters of the two constitutive radicals. We also make empirical predictions regarding the hypomagnetic field effects on planarian regeneration. 
\section{Results}
\subsection{Summary of the weak magnetic field effects on planarian regeneration}
Van Huizen et al.~\cite{van2019weak} reported that WMFs change ROS levels, which in turn alters the expression of heat shock protein 70, which can lead to changes in stem cell proliferation and differentiation, hence regulating blastema formation. In particular, they observed that a WMF affected planarian regeneration in a strength–dependent manner. The planaria were amputated above and below the pharynx cavity, located roughly in the center of the body, and blastemas were allowed to grow under WMF exposure. The WMF was switched on 5 minutes post-amputation. Blastema sizes were analyzed after 3 days of growth. Compared to geomagnetic control (45 $\mu T$), they observed significant reductions in blastema size for field strengths from 100 to 400 $\mu T$ but saw a substantial increase at 500 $\mu T$. The most prominent effects were seen for 200 and 500 $\mu T$ field strength values. They also reported that WMF exposure, to have an effect, was required early and needed to be maintained throughout blastema formation (24-72 hours post-amputation).

\par
As mentioned in the introduction section, ROS plays an essential role in blastema formation, and their concentrations peak at the wound site around 1 to 4 hours post-amputation~\cite{pirotte2015reactive}. Observations of Van Huizen et al.~\cite{van2019weak} also highlight the importance of ROS. They found that pharmacological inhibition of ROS by diphenyleneiodonium resulted in a considerable decrease in blastema sizes. Moreover, they found that by inhibiting superoxide dismutase (SOD), an enzyme that catalyzes \ch{O2^{.-}} removal, they were able to rescue blastema growth in 200 $\mu T$ fields. They also found that SOD inhibition significantly increases blastema sizes in planaria exposed to geomagnetic conditions. Based on this evidence, they hypothesized that WMF effects were mediated by changing ROS concentrations. To confirm this hypothesis, they measured the ROS levels using the cell-permeant fluorescent general oxidative stress indicator dye after 1 hour of amputation for worms exposed to 200 and 500 $\mu T$ fields. As expected, measurements at 200 $\mu T$ revealed a significant decrease in ROS levels, whereas at 500 $\mu T$, they saw increased ROS concentrations.
\par
Although Van Huizen et al.~\cite{van2019weak} concluded that WMF effects are mediated via changes in ROS concentration, they did not identify the specific species involved. \ch{O2^{.-}} and \ch{H2O2} are two of the most biologically important members of the ROS family and are responsible for the majority of ROS signaling. Kinsey et al.~\cite{kinsey2023weak} studied the effects of WMF exposures on these two species during planarian regeneration. They exposed the amputated planaria to 200 and 500 $\mu T$ and then measured \ch{O2^{.-}} and \ch{H2O2} levels using the species-specific dyes after 1 and 2 hours of amputation. They did not observe any significant changes in \ch{H2O2} concentration as compared to geomagnetic control (45 $\mu T$). On the other hand, \ch{O2^{.-}} concentrations were found to be sensitive to WMFs in a fashion similar to WMF effects on ROS-mediated stem cell activity. They reported that \ch{O2^{.-}} concentration decreased for worms exposed to 200 $\mu T$ at both 1 and 2 hours post-amputation. In contrast, no significant change was observed for 500 $\mu T$ fields after 1 hour, and a substantial increase was recorded at 2 hours post-amputation. The WMF effects on \ch{O2^{.-}} levels were significantly greater 2 hours after amputation than 1 hour after. Table 1 shows the approximate percentage change in \ch{O2^{.-}} concentration for both 200 and 500 $\mu T$ fields with respect to geomagnetic control. They also demonstrated that adding exogenous \ch{H2O2} has no rescue effect on WMF inhibition of blastema formation. Based on these findings, Kinsey et al. concluded that WMF effects on planarian regeneration are not mediated via \ch{H2O2}, as they initially suspected, but rather by \ch{O2^{.-}}. They also found that \ch{H2O2} levels peak much before \ch{O2^{.-}} levels. \ch{H2O2} levels peaked around 1 hour after injury and then decreased considerably by the next hour, whereas \ch{O2^{.-}} did not peak until 2 hours post-amputation.

\begin{table}[ht]
\centering
\begin{tabular}{|c|c|c|}
\hline
\multirow{2}{*}{B ($\mu T$)} & \multicolumn{2}{|c|}{\% change in \ch{[O2^{.-}]} compared to control} \\
 & 1 hour post-amputation & 2 hour post-amputation \\
\hline
200 & -10 & -20 \\
\hline
500 & No significant change & 20 \\
\hline
\end{tabular}
\caption{The approximate percentage change in superoxide concentration (\ch{[O2^{.-}]}) for both 200 and 500 $\mu T$ fields with respect to geomagnetic control after 1 and 2 hours of amputation~\cite{kinsey2023weak}.}
\label{tab:superoxide concentration}
\end{table}

\par 
First, we will see whether an RPM-based \ch{O2^{.-}} production scheme can indeed explain the behavior of magnetic fields on planarian regeneration, as suggested by Kinsey et al. For this purpose, we will compare the prediction of our RPM model for \ch{O2^{.-}} yield with the effects observed by Kinsey et al.~\cite{kinsey2023weak} for 200 and 500 $\mu T$ exposures 2 hours post-amputation. We will look for an agreement regarding the shape of WMF effects between our simulations and empirical observation. After such an agreement has been achieved, we will explore the possibility of superoxide amplification to get the effects of the right size. For this purpose, we will make use of a biology-inspired simple chemical kinetic scheme to model superoxide levels and lay out the condition under which such a scheme would be able to reproduce the most salient aspect of Kinsey et al.'s experiment, namely (i) 20 percent change in \ch{O2^{.-}} concentration at 2 hours for both field strengths, (ii) the opposite sign of effect for 200 and 500 $\mu T$, and (iii) significantly stronger effects 2 hours post-amputation than an hour before.  

\subsection{Radical pair mechanism for superoxide production}
\subsubsection{Biological basis of flavin-superoxide radical pair in superoxide production}
Two primary cellular sources of \ch{O2^{.-}} are the mitochondrial electron transport chain and a membrane enzyme family called NADPH oxidase (Nox)~\cite{bedard2007nox,terzi2020role,wallace2010mitochondrial,zhao2019mitochondrial}.
\par
In mitochondria, most of \ch{O2^{.-}} is produced at two sites in complex I and one in complex III. Of the two chemical processes responsible for \ch{O2^{.-}} production in complex I of mitochondria one involves an electron transfer from the reduced form of flavin mononucleotide (\ch{FMNH^{-}}) to molecular oxygen (\ch{O2}) forming \ch{O2^{.-}}  and \ch{FMNH^{.}}~\cite{markevich2015computational}. 
\par
Nox are flavohemoproteins and electron transporters, and Nox1-3 and Nox5 are known to produce \ch{O2^{.-}}. The catalytic core of Nox enzymes is composed of six $\alpha$-helical transmembrane domains. At the C-terminus, the sixth transmembrane domain is linked to a flavin adenosine diphosphate (FAD) binding domain, and the FAD-binding domain is, in turn, linked to an NADPH binding domain. The N-terminus of six transmembrane helices binds to two non-identical heme groups~\cite{terzi2020role}. Nox enzymes transport a hydride (\ch{H^{-}}) anion from NADPH to FAD, resulting in \ch{FADH^{-}}. Electrons from \ch{FADH^{-}} are then transported to \ch{O2}, occupying a binding site near the heme groups~\cite{wu2021mechanistic}. 
\par 
The electron transfers from fully reduced flavin to \ch{O2} during the production of \ch{O2^{.-}} in both mitochondria and Nox, as well as the magnetic field dependence of \ch{O2^{.-}} production strongly suggests the involvement of flavin-superoxide RP (\ch{[FH^{.}... O2^{.-}]}). This underlying \ch{FH^{.}} and \ch{O2^{.-}} based RPM can serve as the basis for explaining various WMF effects observed in the context of for \ch{O2^{.-}} production~\cite{zhang2021long,kinsey2023weak}. However, it should be noted that free \ch{O2^{.-}} is subject to fast spin relaxation, which can severely affect the magnetosensitivity of such an RP. We will address this issue in more detail later in the discussion section.
\par 
At this point, it should also be noted that another flavoprotein called cryptochrome (CRY) is shown to be responsible for blue light-induced \ch{O2^{.-}} production in plants~\cite{el2017blue}. The RPM model for avian magnetoreception is founded on the RPs involving CRY. Tryptophan radical is considered the canonical partner for flavin in CRY RPs. However, an alternative RP with \ch{O2^{.-}} as the partner for flavin can also be formed~\cite{hogben2009possible,muller2011light,lee2014alternative}. This CRY-based \ch{FH^{.}}- \ch{O2^{.-}} RP may be involved in blue light-induced \ch{O2^{.-}} production and, therefore, may render it sensitive to WMFs. However, it is important to mention here that it is unclear whether CRY plays any role in planaria.
\subsubsection{Radical pair mechanism model}

\begin{figure}[ht]
\centering
    \includegraphics[scale=1]{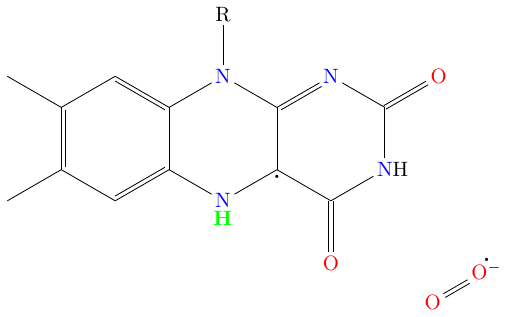}
    \caption{Flavin-superoxide Radical pair}
    \label{Fig:RP}
\end{figure}

\begin{figure}[ht]
\centering
    \includegraphics[scale=1]{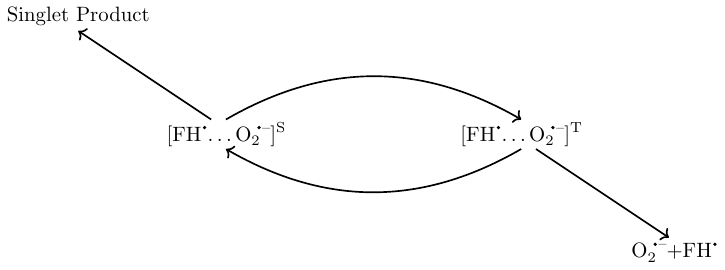}
    \caption{Radical pair reaction scheme}
    \label{Fig:RPM}
\end{figure}

\par
We propose an RP system consisting of \ch{FH^{.}} and \ch{O2^{.-}} (See Fig~\ref{Fig:RP}) to reproduce the non-linear response of magnetic fields on planarian regeneration based on changes in the \ch{O2^{.-}} yields at
200 and 500 $\mu T$ as compared to geomagnetic control. The correlated spins of RP are taken to be in the \ch{[FH^{.}... O2^{.-}]} form. In the triplet state, the radicals diffuse away, releasing fully reduced flavin radical (\ch{FH^{.}}) and \ch{O2^{.-}}. Whereas in singlet state, \ch{FH^{.}} and \ch{O2^{.-}} combine to form some singlet product (See Fig~\ref{Fig:RPM}). Note that we do not identify any particular chemical species as the singlet product. There is also a possibility of \ch{O2^{.-}} being a singlet product rather than a triplet one. This should not be rejected outright since the ground state of the oxygen molecule is a triplet, not a singlet. For concreteness, we have only considered the scenario with \ch{O2^{.-}} as a triplet product.


\par
The most general spin Hamiltonian for RP will include Zeeman ($\hat{H}_{Z}$) and HF ($\hat{H}_{HF}$)
interactions as well as the inter-radical interactions ($\hat{H}_{IR}$), which incorporate exchange and dipolar terms.
\begin{equation}\label{eq:Ham}
\centering
\hat{H} = \hat{H}_{Z}+\hat{H}_{HF}+\hat{H}_{IR}.	
\end{equation}
\par 
The Zeeman term for our RP reads as follows:
\begin{equation}\label{eq:Zee}
\centering
\hat{H}_{Z} = \omega \hat{S}_{{A}_{z}}+\omega \hat{S}_{{B}_{z}}, 	
\end{equation}
where ${\hat{S}}_{A_{z}}$ and ${\hat{S}}_{B_{z}}$ are the spin-z operators of radical electron $A$ (\ch{FH^{.}}) and $B$ (\ch{O2^{.-}}), respectively, and $\omega$ is the Larmor precession frequency of the electrons due to the Zeeman effect. 
\par
Due to the potential random orientation of the molecules in question, we only take into account the isotropic Fermi contact contributions in HF interactions. The unpaired electron on \ch{O2^{.-}}(containing two \textsuperscript{16}O nuclei) has no HF interactions. For the radical electron on \ch{FH^{.}}, for simplicity, we assume in our calculations that unless stated otherwise, the electron couples only with the H5 nucleus. H5 has the largest isotropic HF coupling constant (HFCC) among all the nuclei in \ch{FH^{.}}~\cite{lee2014alternative} and is highlighted in green in Fig~\ref{Fig:RP}. Under these assumptions, the HF term of the Hamiltonian will read as follows:
\begin{equation}\label{eq:HFI}
\centering
\hat{H}_{HF} = a_{1}\mathbf{\hat{S}}_{A}.\mathbf{\hat{I}_{1}}, 	
\end{equation}
where $\mathbf{\hat{S}}_{A}$ is the spin vector operator of radical electron $A$, $\mathbf{\hat{I}_{1}}$ is the nuclear spin vector operator of the H5 of \ch{FH^{.}}, and $a_{1}$ is the isotropic HFCC between the H5 of \ch{FH^{.}} and the radical electron A ($a_{1} = -802.9$ $\mu T$)~\cite{lee2014alternative}.
\par
Furthermore, for simplicity, we do not consider any inter-radical interactions in our model except when explicitly stated otherwise.
\par 
The singlet and triplet reaction rates are denoted by $k_S$ and $k_T$, respectively. The spin relaxation rates of radicals A and B are denoted by $r_A$ and $r_B$, respectively.
\par
As mentioned earlier \ch{[FH^{.}... O2^{.-}]} RP can start as a triplet, singlet, or F-pair. In this study, we have all these initial states. We have also considered the effects of the exchange interaction and multiple HF interactions for \ch{FH^{.}} in the supporting information.

\subsubsection{Triplet-born radical pair}
Since the ground state of the oxygen molecule is a triplet, let us start by considering the initial state of the RP to be a triplet:
\begin{equation}\label{eq:triplet}
	 \frac{1}{3M}\hat{P}^T=\frac{1}{3}\Big{\{}\ket{T_{0}}\bra{T_{0}}+\ket{T_{+1}}\bra{T_{+1}}+\ket{T_{-1}}\bra{T_{-1}}\Big{\}}\otimes \frac{1}{M} I_{M},
	\end{equation}
where $\hat{P}^T$ is the triplet projection operator, $M$ is the total number of nuclear spin configurations, $\ket{T_{0}}$ and $\ket{T_{\pm1}}$ represent the triplet states of two electrons in RP with the spin magnetic quantum number ($m_{S}$) equal to $0$ and $\pm1$ respectively. $\hat{I}_{M}$ represents the completely mixed  initial state of the nuclei.
\par
The fractional triplet yield generated by the RPM can be determined by monitoring the dynamics of RP spin states. For details of the calculations, see the Methods section. The ultimate fractional triplet yield ($\Phi^{(T)}_{T}$) for a RP that originates in a triplet state, when considering time intervals significantly longer than the RP's lifetime, is as follows:
\begin{equation}\label{eq:triplet born triplet yield}
\centering
\Phi^{(T)}_{T} = k_{T}\text{ }\text{Tr}\Big[\hat{P}^{T}\hat{\hat{L}}^{-1}[\frac{1}{3M}\hat{P}^T]\Big],
\end{equation}
where $\hat{\hat{L}}$ is the Liouvillian superoperator and other symbols have above stated meanings.
\par

There are four free parameters in our model, namely, $k_S$, $k_T$, $r_A$, and $r_B$. Our task is to figure out regions in our parameter space where simulated behavior corresponds to the experimental observation (i.e., a positive change in $\Phi^{(T)}_{T}$ at 500 $\mu T$  and a negative change at 200 $\mu T$ with respect to geomagnetic control). From now onward, such a region will be referred to as the allowed region. For this purpose, we investigated the signs of triplet yield changes with respect to control at 200 $\mu T$ and 500 $\mu T$ over a wide range of chemical reaction rates ($k_S \in \{10^{4} $ $s^{-1},10^{7}$ $ s^{-1}\} $ and $k_T \in \{10^{4}$ $ s^{-1},10^{7}$ $ s^{-1}\}$) for various pairs of spin relaxation rates $r_A$ and $r_B$ (See the supporting information). We observed that such an allowed region in $k_S$-$k_T$ plane can be found provided $r_A\leq10^5$ $ s^{-1}$ and $r_B\leq10^6$ $ s^{-1}$.
\par
\begin{figure}[ht]
     \centering
     \begin{subfigure}[b]{0.8\textwidth}
         \centering
         \includegraphics[width=\textwidth]{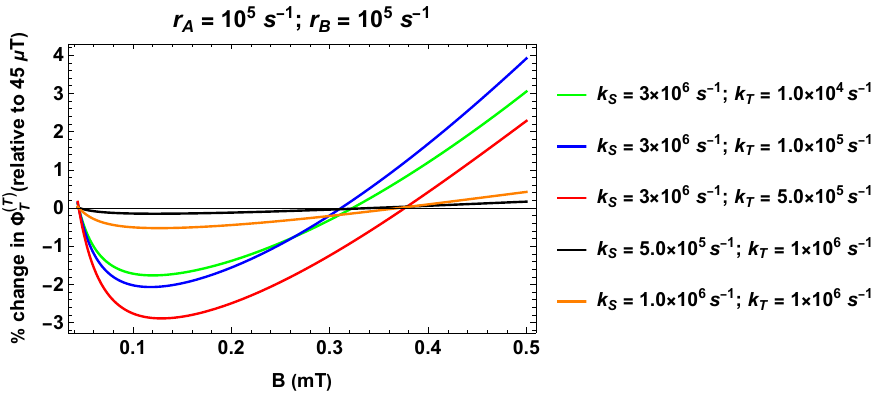}
         \caption{}
         \label{Fig:Triplet2a}
     \end{subfigure}
     \hfill
     \begin{subfigure}[b]{0.8\textwidth}
         \centering
         \includegraphics[width=\textwidth]{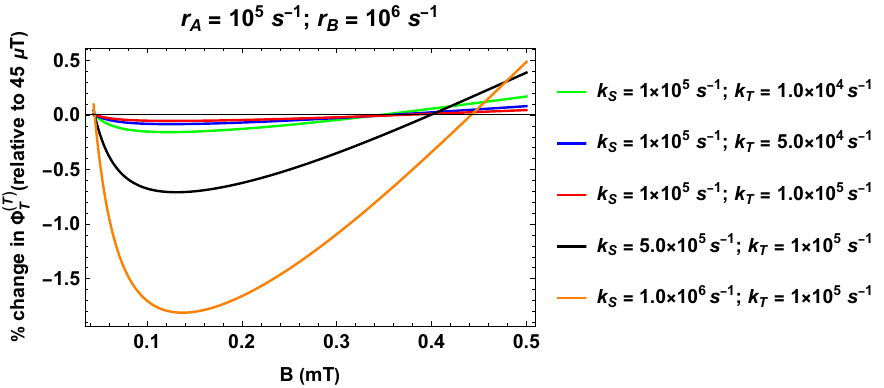}
         \caption{}
         \label{Fig:Triplet2b}
     \end{subfigure}
      \caption{Percentage change in the fractional triplet yield for triplet-born RP with respect to the geomagnetic control (45 $\mu T$) as a function of the magnetic field. The HFCC value $a_1$ is taken to be $802.9$ $\mu T$. (a) $r_A=10^5$ $s^{-1}$ and $r_B=10^5$ $s^{-1}$. (b) $r_A=10^5$ $s^{-1}$ and $r_B=10^6$ $s^{-1}$. $k_S$ and $k_T$ are singlet and triplet reaction rates, respectively. $r_A$ and $r_B$ are the spin relaxation rates of radicals A and B, respectively.}
        \label{Fig:Triplet2}
\end{figure}

 
Fig.~\ref{Fig:Triplet2} shows the percentage change in $\Phi^{(T)}_{T}$ with respect to the geomagnetic control as a function of the magnetic field for various allowed values of $k_S$ and $k_T$ when $r_A$ and $r_B$ are both fixed. In Fig.~\ref{Fig:Triplet2a} both $r_A$ and $r_B$ values are fixed at $1\times10^5$ $s^{-1}$. In Fig.~\ref{Fig:Triplet2b} $r_A$ is again fixed at $1\times10^5$ $s^{-1}$ and $r_B$ is fixed  to $1\times10^6$ $s^{-1}$. These figures show that larger $k_S$ correspond to larger percentage changes at  200 $\mu T$ and 500 $\mu T$. For $k_S=3\times10^6$ $s^{-1}$ in Fig~\ref{Fig:Triplet2a}, we can have changes of about $2-3\%$. An effect size of about $1\%$ can be reached for $k_S=1\times10^6$ $s^{-1}$ in Fig.~\ref{Fig:Triplet2b}. It should be noted that lowering the $r_A$ value had no significant effect on any of these effects (see the supporting information).

\subsubsection{Singlet-born radical pair}
Now, we will consider the initial state of the RP to be a singlet:
\begin{equation}\label{eq:singlet}
	 \frac{1}{M}\hat{P}^S= \ket{S}\bra{S}\otimes \frac{1}{M}I_{M},
	\end{equation}
where  $\hat{P}^S$ is the singlet projection operator and $\ket{S}$ represents the singlet state of two electrons in RP.
The ultimate fractional triplet yield ($\Phi^{(S)}_{T}$) for a RP that originates in a singlet state for time intervals significantly longer than the RP's lifetime is given as:
\begin{equation}\label{eq:singlet born triplet yield}
\centering
\Phi^{(S)}_{T} = k_{T}\text{ }\text{Tr}\Big[\hat{P}^{T}\hat{\hat{L}}^{-1}[\frac{1}{M}\hat{P}^S]\Big].
\end{equation}
As for triplet-born RP, we investigated the signs of triplet yield changes with respect to control at 200 $\mu T$ and 500 $\mu T$ over a wide range of chemical reaction rates ($k_S \in \{10^{4}$ $ s^{-1},10^{7}$ $ s^{-1}\} $ and $k_T \in \{10^{4}$ $ s^{-1},10^{7}$ $ s^{-1}\}$) for various pairs of spin relaxation rates ($r_A$ and $r_B$). We could not find any region in the $k_S$-$k_T$ plane for which $\Phi^{(S)}_{T}$ changes in accordance with experimental observations (See the supporting information). Therefore, a singlet-born RP is ruled out as an explanation for \ch{O2^{.-}} mediated WMF effects on planarian regeneration. Note, singlet-born RP could have been able to explain the empirical observations if \ch{O2^{.-}} were to be a singlet product rather than a triplet one in the RP reaction scheme (Fig.~\ref{Fig:RPM}).

\subsubsection{F-pair radical pair}
Apart from singlet and triplet states, RPs can start as F-pairs. These are formed by random encounters of the two constitutive radicals. F-pair state is a completely mixed state and is given as $\frac{1}{4M}\hat{I}_{4M}$, where $\hat{I}_{4M}$ is an identity matrix of dimension $4M$. Such RPs behave similarly to triplet-born pairs in the presence of an effective singlet channel (i.e., $k_S>k_T$) and behave similarly to singlet-born pairs in the presence of effective triplet channel (i.e., $k_S<k_T$)~\cite{kattnig2021f}. It should also be pointed out that for allowed regions, the size of WMF effects is similar in strength to that of triplet-born RPs in the corresponding regions (See the supporting information).

\subsubsection{Hypomagnetic effects}
\begin{figure}[ht]

          \centering
     \begin{subfigure}[b]{0.8\textwidth}
         \centering
         \includegraphics[width=\textwidth]{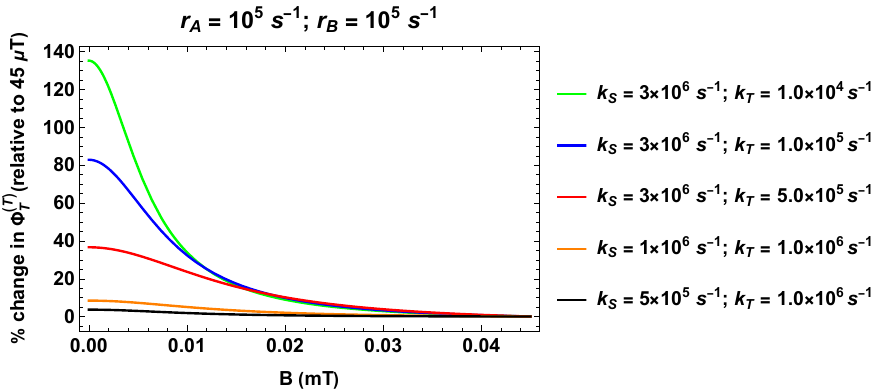}
         \caption{}
         \label{Fig:HMFa}
     \end{subfigure}
     \hfill
     \begin{subfigure}[b]{0.8\textwidth}
         \centering
         \includegraphics[width=\textwidth]{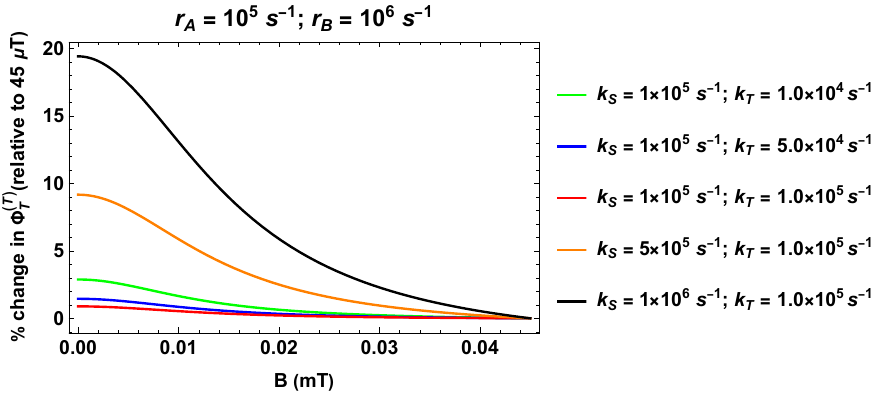}
         \caption{}
         \label{Fig:HMFb}
     \end{subfigure}

      \caption{ Percentage change in the fractional triplet yield for triplet-born RP with respect to the geomagnetic control (45 $\mu T$) as a function of the magnetic field. (a) $r_A=10^5$ $ s^{-1}$ and $r_B=10^5$ $ s^{-1}$. (b) $r_A=10^5$ $ s^{-1}$ and $r_B=10^6$ $ s^{-1}$. $k_S$ and $k_T$ are singlet and triplet reaction rates, respectively. $r_A$ and $r_B$ are the spin relaxation rates of radicals A and B, respectively.}
       \label{Fig:HMF1}

\end{figure}

In an RPM, the fractional triplet yield can also be altered by shielding the geomagnetic field~\cite{zadeh2023hypomagnetic}. The effect on the fractional triplet yield of the RPM of shielding geomagnetic field for a triplet-born RP is shown in Fig~\ref{Fig:HMF1}. Our simulation suggests a positive change similar to $500$ $\mu T$ field. However, it should be noted that the exact size of the effect will depend on the various rate values.

\subsection{Amplification of magnetic field effects on reactive oxygen yield}
As we have seen so far, RPM can predict WMF effects at 200 and 500 $\mu T$ up to a few percentage points for a suitable choice of rates. On the other hand, the percentage changes 2 hours post-amputation observed by Kinsey et al.~\cite{kinsey2023weak} are around 20 percent. Despite its ability to predict the correct behavior of WMF effects, RPM alone can not produce effects of the right size and does not explain the temporal pattern observed, i.e., changes observed are much more pronounced after two hours than after one hour of amputation. These observations suggest the need for an amplification pathway.
\par
The possibility of such an amplification scheme has been suggested several times in the biological literature in the past. The existence of a \ch{Ca^{2+}}-ROS self-amplifying loop has been observed in plants~\cite{pottosin2018powering}. Nox enzyme is activated by cytosolic \ch{Ca^{2+}}, and in turn, the superoxide produced activates \ch{Ca^{2+}} influx across the plasma membrane. Another superoxide amplification loop relies on the JNK signaling pathway~\cite{chambers2011mitochondrial}. ROS activates the JNK pathway, and mitochondrial JNK signaling is, in turn, responsible for mitochondrial superoxide amplification. This pathway is known to be activated during blastema formation in planaria~\cite{tasaki2011role}.
Moreover, there is evidence of Nox-mitochondria cross-talk~\cite{dikalov2011cross}. Mitochondria act as a target for ROS produced by Nox, and under certain conditions, mitochondrial ROS may stimulate NADPH oxidases. Therefore, this cross-talk between mitochondria and NADPH oxidases may represent a feed-forward cycle of ROS production.
\par
Inspired by these positive feedback loops, we propose a simple chemical kinetic scheme to model superoxide amplification. 
\begin{equation}\label{eq:amplification}
\centering
\frac{d\ch{[O2^{.-}]}}{dt} = k_1\times\Phi^{(T)}_T + k_2\times\ch{[O2^{.-}]}^2 - k_3\ch{[O2^{.-}]}.
\end{equation}
The first term represents superoxide production and is assumed to proceed via RPM, as discussed in the preceding sections. Following Player et al. ~\cite{player2021amplification}, we can model the effect of applied magnetic fields on the kinetics of the superoxide production step simply by multiplying the overall reaction constant by $\Phi^{(T)}_T $. The last term models the superoxide removal. The middle term is the phenomenological self-amplification term. The dependence on superoxide must be at least second-degree for any amplification of magnetic field effects to happen. We have taken the simplest scenario.
\par
Kussmaul et al. have stated that under ``normal" conditions (respiration supported by malate/pyruvate), superoxide production by complex I of mitochondria is very low ($< 0.05$ $nmol$ $min^{-1}mg^{-1}$ $\approx10^{-6} $ $M s^{-1}$)~\cite{kussmaul2006mechanism}. The unit conversion is made assuming water to be the solvent. It should also be noted that the superoxide is primarily removed by SOD. The rate of disproportionation of superoxide catalyzed by SOD is of the order of $10^{9}$ $ M^{-1} s^{-1}$~\cite{sheng2014superoxide,mailloux2015teaching}. For these values, our model does not produce any amplification of magnetic field effects for any reasonable values of $k_2$ unless SOD concentrations are very low ($\sim 10^{-7}$ $ \mu M$). However, it is unlikely that the rates for amputated planaria would be similar to these ``normal condition" values.   
\par
Equation~\ref{eq:amplification} is a nonlinear differential equation, and we used phase space analysis to infer some of the features of its possible solutions. Depending upon the initial \ch{O2^{.-}} concentration and the values of various chemical rates involved, we can either get solutions where \ch{O2^{.-}} concentration saturates to an equilibrium value (saturating solution), or it keeps increasing (amplifying solution). If the discriminant of the quadratic polynomial on the right-hand side of Eq.~\ref{eq:amplification} is negative (i.e., ${k_{3}}^2-4k_{2}k_{1}\Phi^{(T)}_T<0$) then there are no real roots of this polynomial and the differential equation will always produce amplifying solutions irrespective of initial concentration of  \ch{O2^{.-}}. If the discriminant is non-negative, then the polynomial has two real roots, and we can get either saturating or amplifying solution depending on the initial condition. To get amplification in this case, the initial concentration of  \ch{O2^{.-}} must be greater than the larger of the two roots. A much more stringent restriction on our model's allowed values of chemical rates is put by the condition that the size of magnetic field effects must reach around 20\% at the two hour mark. 
\par
We analyzed the solutions of Eq.~\ref{eq:amplification} for a large range of chemical rate values. We started by fixing the values of $k_1$ and $k_3$ and then looking for the values of $k_2$ for which magnetic field effects reach around 20\% at the two-hour mark. We did this process for various reasonable values of $k_1$ and $k_3$. We found that for our model to work $k_1$ should be of the order of $10^{-5} s^{-1}$ or smaller and $k_3$ should be of the order of $10^{-3}$ $s^{-1}$ or smaller.  
\par
\begin{figure}[ht]
     \centering

         \includegraphics[width=1.0\textwidth]{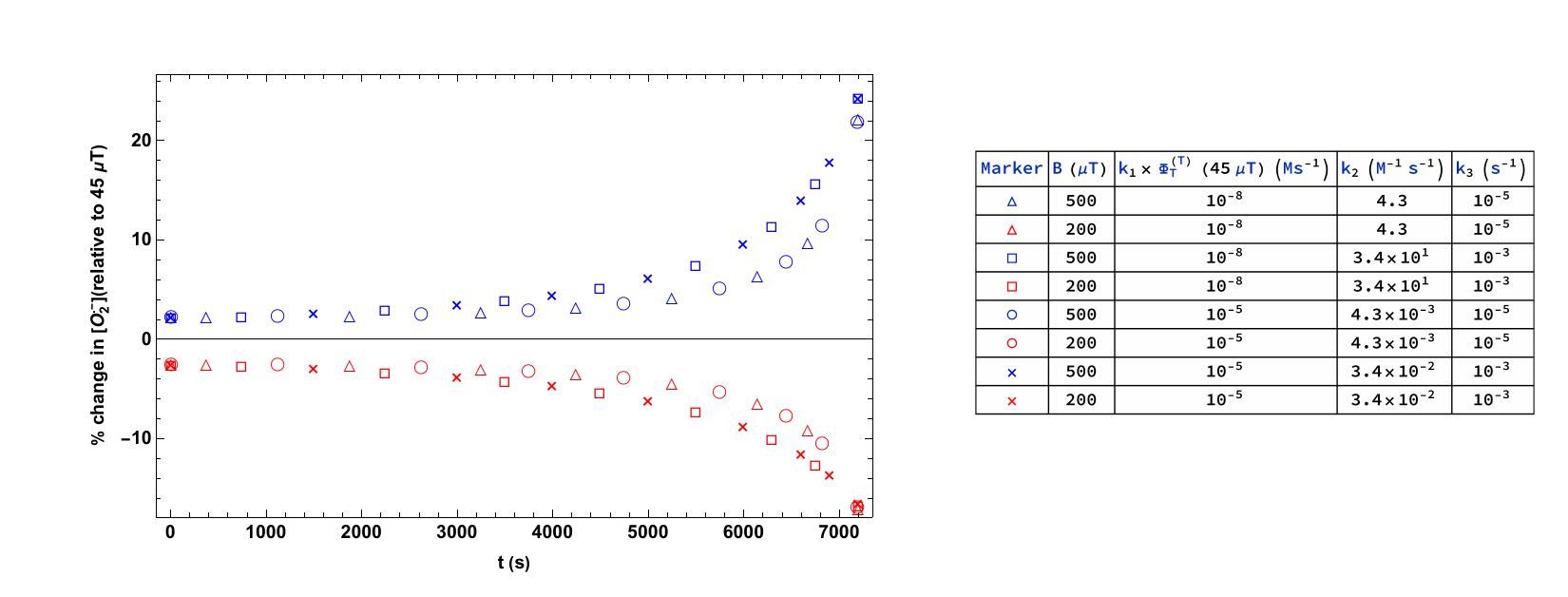}

      \caption{Percentage change in the superoxide concentration with respect to the geomagnetic control (45 $\mu T$) for the first two hours after amputation. The strength of the WMF effect at both $200 \text{ and } 500$ $\mu T$ is taken to be $-2.5\%$ and $+2.3\%$ respectively, corresponding to $r_A=10^5$ $ s^{-1}$, $r_B=10^5$ $ s^{-1}$, $k_S=3.0\times10^6$ $s^{-1}$, and $k_T=5.0\times10^5$ $s^{-1}$. $k_1$, $k_2$, and $k_3$ are defined in Eq.~\ref{eq:amplification}. $k_S$ and $k_T$ are singlet and triplet reaction rates, respectively. $r_A$ and $r_B$ are the spin relaxation rates of radicals A and B, respectively.}
       \label{Fig:Amplification1}
    
\end{figure}

With suitable values for the chemical rates, this simple amplification model can reproduce the most salient aspect of Kinsey et al.’s experiment~\cite{kinsey2023weak}. Fig.~\ref{Fig:Amplification1} shows the percentage change in the superoxide concentration with respect to the geomagnetic control for the first two hours after amputation for some of these suitable chemical rate values. The strength of the WMF effect at both $200 \text{ and } 500$ $\mu T$ is taken to be $-2.5\%$ and $+2.3\%$ respectively, corresponding to $r_A=10^5$ $ s^{-1}$, $r_B=10^5$ $ s^{-1}$, $k_S=3.0\times10^6$ $s^{-1}$, and $k_T=5.0\times10^5$ $s^{-1}$. As can be seen, it is possible to get a 20 percent change for both field values two hours post-amputation. Moreover, the percentage change at the hour mark is much smaller than the two-hour mark. However, it should be noted that Kinsey et al.~\cite{kinsey2023weak} observed around a 10 percent decrease at the hour mark for $200$ $\mu T$, which this simple model does not capture, instead predicting a substantially smaller change. In reality, the situation is more complex than what our model describes. There are possibly multiple  \ch{O2^{.-}} production, removal, and amplification pathways operating at the same time. It should be noted that the concentration of \ch{O2^{.-}} after 2 hours will also depend on the chemical rate values. For chemical rate values in Fig.~\ref{Fig:Amplification1}, it is of the order of $10^{-4}$ $M$ (for `$\triangle$'), $10^{-5}$ $M$ (for `$\Box$'), $10^{-1}$ $M$ (for `${\Circle}$'), and $10^{-2}$ $M$ (for `$\times$'). In Fig.~\ref{Fig:Amplification1}, we have taken the initial concentration of  \ch{O2^{.-}} to be $10^{-10}$ M. Similar results can be obtained for other reasonable values of initial concentration.

\section{Discussion}
In this paper, we set out to analyze Kinsey et al.'s~\cite{kinsey2023weak} suggestion that the RPM can explain their observations regarding WMF effects on ROS levels during planarian regeneration. Our results point to the possibility of such a mechanism. We found that the RPM model for a suitable choice of parameters can reproduce the non-linear dependence on WMFs observed in the experiment at a level of \ch{O2^{.-}} concentration. It is important to note that we have not modeled the regeneration process itself in this study and that this would be an interesting direction for future research. 
\par 
We propose that a \ch{[FH^{.}... O2^{.-}]} RP-based model similar to that used in the previous studies on WMF effects on ROS~\cite{usselman2016quantum,rishabh2022radical} may be responsible for the non-linear dependence on WMFs on \ch{O2^{.-}} concentration, which in turn results in similar behavior at the level of blastema size. A similar RP has been proposed in the context of avian magnetoreception~\cite{hogben2009possible,muller2011light,lee2014alternative}. The magnetic field effects on the circadian clock and isotopic dependence of behavioral effects of Lithium have also been explained based on \ch{[FH^{.}... O2^{.-}]} RP~\cite{zadeh2022Circadian,zadeh2021Li}.
\par
The results of our calculations favor triplet-born RP rather than a singlet-born RP. Mostly in RPs involving \ch{O2^{.-}}  and \ch{FH^{.}}, the initial state of RP is taken to be a triplet state. It is because, in the RP formation process, the oxygen molecule is in its triplet ground state before the electron transfer from flavin. Consequently, the initial state of the RP formed would be a triplet state. However, our previous study on the RPM-based model for hypomagnetic field effects on neurogenesis found that singlet instead of triplet-born RP agreed with observations on superoxide concentration~\cite{rishabh2022radical}. There also we considered \ch{O2^{.-}} to be a triplet product. If we considered it a singlet product, triplet instead of singlet-born RP would have agreed with the observations. These differences between the neurogenesis and regeneration might suggest the involvement of different RPs in the two cases. Given that multiple superoxide sources (including mitochondria and Nox) are present in a cell, this may not be an implausible scenario. It should also be pointed out that F-pairs, which are formed by random encounters of the two constitutive radicals, can also explain the empirical observation in the presence of an effective singlet channel.
\par
Unlike previous studies~\cite{usselman2016quantum,rishabh2022radical}, we do not assume \ch{H2O2} to be the singlet product as this would contradict the empirical observations. Kinsey et al.~\cite{kinsey2023weak} did not observe any significant effect of WMF on \ch{H2O2} concentration. 
\par
It has been suggested in the past that due to fast molecular rotation, free \ch{O2^{.-}} has a spin relaxation lifetime on the orders of $1$ $ns$ and hence a fast spin relaxation rate $r_B$~\cite{hogben2009possible,player2019viability}. The relaxation rate requirement calculated by our model yields $r_B$ at least 3 order lower than this expected value. However, this fast spin relaxation of free superoxide can be lowered if the molecular symmetry is reduced and the angular momentum is quenched by the biological environment~\cite{hogben2009possible,player2019viability}. Although, it should be noted that for this to happen \ch{O2^{.-}} must be tightly bound~\cite{player2019viability}. Such a possibility may arise in the case of the Nox enzyme because of the presence of \ch{O2} binding pockets near the heme proteins. Recently, it has been predicted that \ch{O2} can localize at certain sites within the cryptochrome protein for tens of nanoseconds, and \ch{O2^{.-}} can localize for significantly longer~\cite{salerno2023long}. However, it remains unclear whether CRY plays any role in planaria. It has also been indicated that \ch{O2} would need to bind in the mitochondrial electron transfer flavoprotein for superoxide production~\cite{husen2019molecular}. Direct evidence of such inhibition of spin relaxation (for example, an electron paramagnetic resonance spectrum of \ch{O2^{.-}}) has yet to be found. 
\par
For the most part, in our modeling, we have considered only one nuclear spin. It should be noted that an additional nuclear spin shifts the allowed region to slightly higher rate values and introduces wiggles around 150 $\mu T$ (See the supporting information). No major change is observed regarding the behavior of WMF effects at 200 and 500 $\mu T$ in the allowed region. Thus, having multiple HF interactions does not change our main conclusions.
\par
Exchange and dipolar interactions can significantly affect the response of a RP to an applied magnetic field. No dipolar interaction has been included in our model. We have considered the effects of exchange interaction in the supporting information. It is generally expected that the radicals should be far apart for these interactions to be neglected. However, Efimova and Hore~\cite{efimova2008role} have shown that at the small separation of $2.0 \pm 0.2$ nm, the effects of exchange and dipolar interactions can partially cancel. It should also be noted that electron transfer via an electron transfer chain could eliminate the need for two radicals to be close enough to have any significant inter-radical interactions.
\par
We also found that tightly bound flavin molecules, for which we need to consider anisotropic rather than isotropic HF coupling, can not explain Kinsey et al.'s observation (See the supporting information). This suggests that either the \ch{O2^{.-}} involved is not produced via Nox, or the flavin bound to Nox is still free to rotate.
\par
Our RPM-based model for WMF effects also points to a hypomagnetic field effect on planarian neurogenesis. Our simulation suggests a positive change with respect to geomagnetic control. For some rate values, this change can be much stronger than applying the 500 $\mu T$ field. It would be interesting to conduct experiments to explore the impact of shielding the Earth's magnetic field. Further, it would also be interesting to see larger fields' effect on \ch{O2^{.-}} concentration. Beane's group has measured the effect of such a field on blastema size and general ROS but not on \ch{O2^{.-}}. Their observation shows no significant effect on blastema size for fields ranging from 500-900 $\mu T$. If these patterns are emulated by \ch{O2^{.-}} levels, then they would challenge the present simple RPM and superoxide amplification model. This might indicate that some deamplification mechanism is switched on (or the amplification mechanism is switched off) when \ch{O2^{.-}} levels get too large. Applying an oscillating external magnetic field is also an interesting avenue for future research.
\par
It would be interesting to check our predictions regarding the substitution of one \ch{^{16}O} with \ch{^{17}O} in the superoxide radical. We found no significant change in the strength of effects at $200$ $\mu T$, whereas the effects at $500$ $\mu T$ were slightly reduced (See the supporting information). It should be noted that the exact nature of the effect will be a function of precise parameter values. Experiments involving the substitution of \ch{^{16}O} with \ch{^{17}O} are familiar to researchers. Several such experiments have been performed in different biological contexts~\cite{paech2020quantitative,fiat1992vivo,mellon2009estimation}. Some of these experiments~\cite{mellon2009estimation,paech2020quantitative} have been performed with up to $70\%$ \ch{^{17}O} enrichment, making our assumption of one \ch{^{17}O} HFI reasonable.
\par
Kattnig et al. have proposed that fast spin relaxation of the \ch{O2^{.-}} can be dealt with by involving a third scavenger radical~\cite{kattnig2017radical,kattnig2017sensitivity, ramsay2022radical}. Babcock et al.~\cite{babcock2021radical} have also suggested that radical scavenging could resolve the challenge posed by electron-electron dipolar interactions. Given a suitable choice of scavenger radical, we have shown that such a radical triad model can explain the observed effects (See the supporting information). Because of the fast spin relaxation of superoxide, no magnetic effect is expected for the substitution of \ch{^{16}O} with \ch{^{17}O}~\cite{ramsay2022radical} in this scenario.
\par
Despite predicting the correct behavior of magnetic field effects, the RPM model alone can not predict the right size of these effects and does not account for the temporal aspect of Kinsey et al.'s~\cite{kinsey2023weak} observation. We proposed a simple chemical kinetic scheme to model \ch{O2^{.-}} concentration. This model included a magnetic field-dependent production term, a \ch{O2^{.-}} removal term, and a positive feedback term to model amplification. The existence of \ch{Ca^{2+}}-\ch{O2^{.-}} self-amplifying loop~\cite{pottosin2018powering} and JNK-\ch{O2^{.-}} amplification pathway~\cite{chambers2011mitochondrial}, and the fact that such a pathway is activated precisely during regeneration~\cite{dikalov2011cross} adds to the plausibility of such an amplification term in our model. Inhibition experiments can be an interesting test for the role played by these amplification mechanisms. For example, if JNK signaling inhibition results in much decreased ROS levels and, hence, reduced regeneration, then this would strongly indicate the involvement of JNK-\ch{O2^{.-}} amplification pathway. Similar experiments for  \ch{Ca^{2+}}-\ch{O2^{.-}} self-amplifying loop could also be performed by blocking \ch{Ca^{2+}}-channels. We found that for this model to work, the \ch{O2^{.-}} removal mechanism must be significantly inefficient. It should be noted that this model predicts a smaller effect at the hour mark for 200 $\mu T$ than that observed. In reality, the situation is more complex than what our model describes. There are possibly multiple  \ch{O2^{.-}} production, removal, and amplification pathways operating at the same time. While our model is admittedly simple, it suggests that the RPM as a potential underlying mechanism for the WMF effects on planarian regeneration should be taken seriously while also pointing to the need for an amplification pathway.

\par
 Let us also note that the underlying RP might not have flavin as the partner for superoxide~\cite{smith2021radical,zadeh2022radical, bradlaugh2023essential, mccormick1978near,jin1993superoxide,hunter1989effect,houee2015exploring}. These WMF effects may also be due to some non-superoxide RP and the changes in \ch{O2^{.-}} concentration, in that case, is due to some back-reaction, as might be the case for CRY~\cite{el2017blue}. It is also possible that explanations other than the RPM may underlie the WMF effects on planarians.
\par
It will be interesting to see whether WMF effects can be observed in other contexts where ROS play a crucial role, such as biophoton emission~\cite{pospivsil2014role}, and whether a similar RPM could explain these effects.

\section{Methods}
\subsection{Radical pair mechanism calculations}
The state of the RP was described using the spin density operator. As shown in the results section, the spin Hamiltonian involve the Zeeman and the HF interactions and interradical interactions. The coherent spin dynamics, chemical reactivity, and spin relaxation all together determine the time evolution of the spin density matrix of the RP system.
\par
The time dependence of the spin density operator was obtained using the Liouville Master Equation~\cite{haberkorn1976density,steiner1989magnetic}:

\begin{equation}\label{eq:ms}
\centering
\frac{d\hat{\rho(t)}}{dt} = -\hat{\hat{L}}[\rho(t)],
\end{equation}
where Liouvillian superoperator $\hat{\hat{L}}=\iota\hat{\hat{H}}+\hat{\hat{K}}+\hat{\hat{R}}$. $\hat{\hat{H}}$, $\hat{\hat{K}}$, and $\hat{\hat{R}}$ are Hamiltonian superoperator (see Eq.~\ref{eq:Ham}), chemical reaction superoperator, and spin relaxation superoperator, respectively. For spin-selective chemical reactions (reaction scheme of Fig.~\ref{Fig:RPM}), we use the Haberkorn superoperator~\cite{haberkorn1976density}, which is given by the following equation:
\begin{equation}\label{eq:K}
\centering
\hat{\hat{K}} = \frac{1}{2}k_{S}\Big(\hat{P}^{S}\otimes I_{4M} + I_{4M} \otimes\hat{P}^{S}\Big) + \frac{1}{2}k_{T}\Big(\hat{P}^{T}\otimes I_{4M} + I_{4M} \otimes\hat{P}^{T}\Big) ,
\end{equation}
where symbols have above stated meanings. Spin relaxation is modeled via random time-dependent local fields~\cite{kattnig2016electron,player2020source}, and the corresponding superoperator reads as follows: 
\begin{equation}\label{eq:R}
\centering
\begin{aligned}
    \hat{\hat{R}} &= r_{A}\Big[\frac{3}{4} I_{4M}\otimes I_{4M} - \hat{S}_{{A}_{x}}\otimes(\hat{S}_{{A}_{x}})^{T}-\hat{S}_{{A}_{y}}\otimes(\hat{S}_{{A}_{y}})^{T}-\hat{S}_{{A}_{z}}\otimes(\hat{S}_{{A}_{z}})^{T}\Big]\\&+r_{B}\Big[\frac{3}{4} I_{4M}\otimes I_{4M} - \hat{S}_{{B}_{x}}\otimes(\hat{S}_{{B}_{x}})^{T}-\hat{S}_{{B}_{y}}\otimes(\hat{S}_{{B}_{y}})^{T}-\hat{S}_{{B}_{z}}\otimes(\hat{S}_{{B}_{z}})^{T}\Big], 
\end{aligned}
\end{equation}
where the symbols have above stated meanings. The ultimate fractional \ch{O2^{.-}} yield for triplet-born RP ($\Phi^{(T)}_T$) for time periods much greater than the RP lifetime is given by:
\begin{equation}\label{eq:triplet yield}
\centering
\Phi^{(T)}_T = k_{T}\text{ }\text{Tr}\Big[\hat{P}^{T}\hat{\hat{L}}^{-1}[\frac{1}{3M}\hat{P}^T]\Big].
\end{equation}
The computational calculations and plotting were performed on Mathematica~\cite{Mathematica}.
\subsection{Radical triad calculations}
Radical triad calculations were carried out in a similar fashion to RPM (see Ref.~\cite{ramsay2022radical} for details). We assumed instantaneous relaxation of superoxide, and no relaxation was assumed for the other two radicals. Further, an effective oxidation process was also assumed.

\section*{Acknowledgement}
The authors thank Dr. Luke Kinsey, Dr. Wendy Beane, and Prof. Dennis Salahub for helpful discussions.
\par
This work was supported by the Natural Sciences and Engineering Research Council through its Discovery Grant and CREATE programs as well as the Alliance Quantum Consortia Grant `Quantum Enhanced Sensing and Imaging', and by the National Research Council of Canada through its Quantum Sensing Challenge Program. 
\section*{Author contributions}
C.S. conceived the project; R. performed the modelling and calculations with help from H.ZH. and C.S.; R. wrote the paper with feedback from H.ZH. and C.S.

\printbibliography

\section*{Supporting Information}

\subsection*{Triplet-born radical pair}
\begin{figure}[ht!]
     \centering

     \begin{subfigure}[b]{0.45\textwidth}
         \centering
         \includegraphics[width=\textwidth]{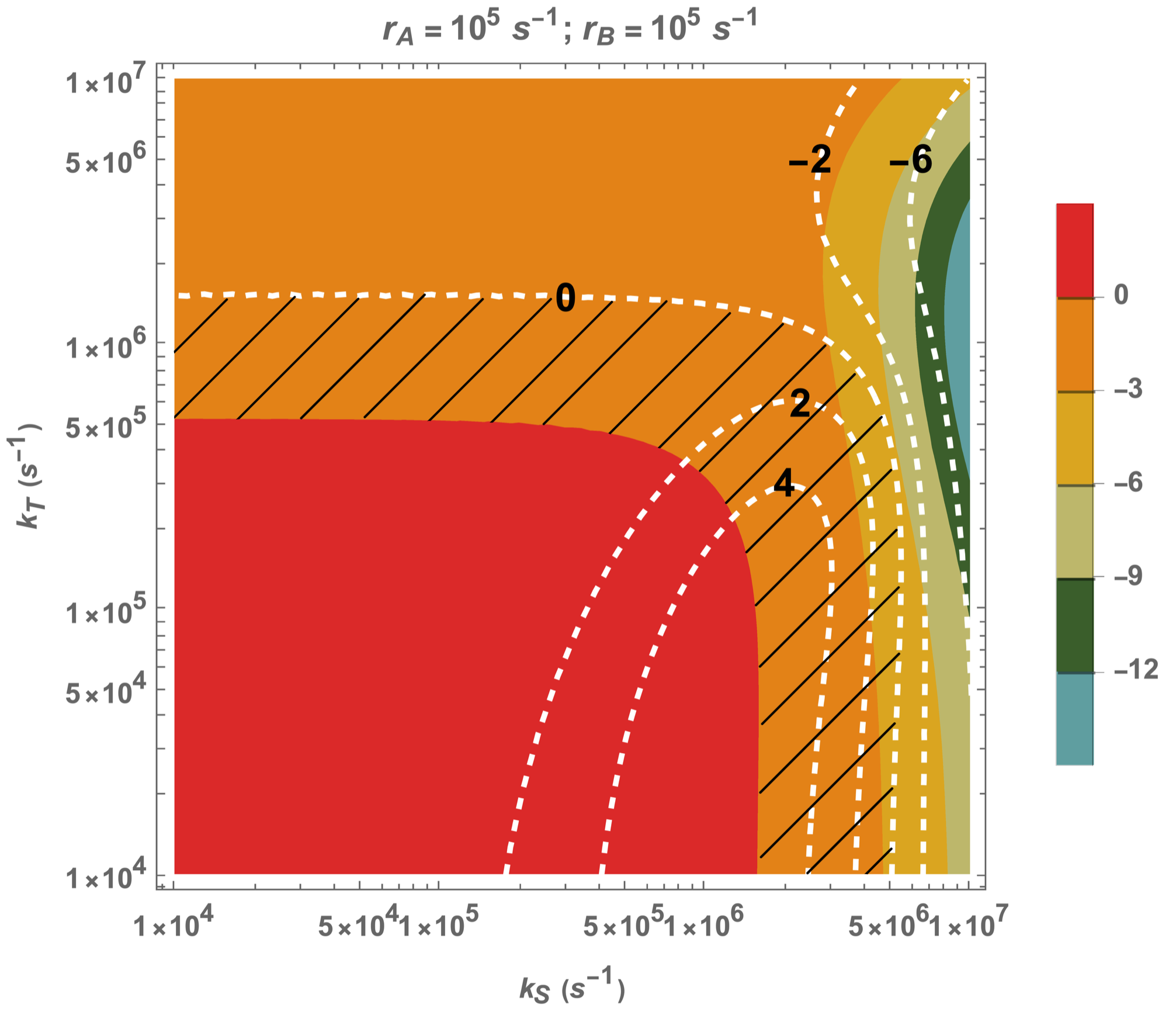}
         \caption{}
         \label{Fig:Triplet1a}
     \end{subfigure}
     \hfill
     \begin{subfigure}[b]{0.45\textwidth}
         \centering
         \includegraphics[width=\textwidth]{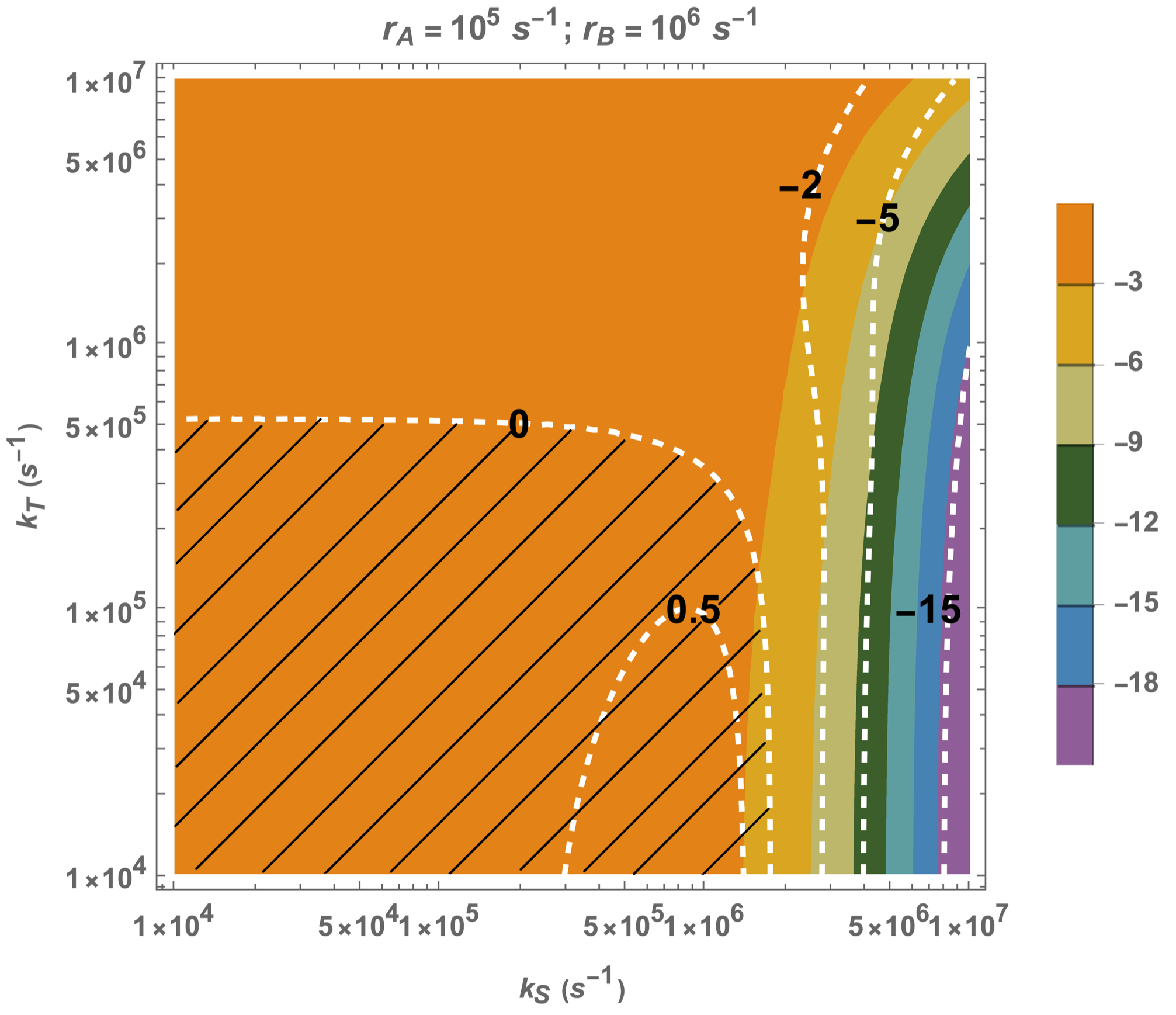}
         \caption{}
         \label{Fig:Triplet1b}
     \end{subfigure}
    
      \caption{The triplet yield changes for triplet-born RP with respect to control (45 $\mu T$) at 200 $\mu T$ and 500 $\mu T$ in $k_{S}-k_{T}$ plane. The colored shading represents the percentage changes in triplet yield at 200 $\mu T$ with respect to geomagnetic control, and the white contour lines represent the percentage changes in triplet yield at 500 $\mu T$ with respect to geomagnetic control. The HFCC value $a_1$ is taken to be $802.9$ $\mu T$. (a) $r_A=10^5$ $s^{-1}$ and $r_B=10^5$ $s^{-1}$. (b) $r_A=10^5$ $s^{-1}$ and $r_B=10^6$ $s^{-1}$. $k_S$ and $k_T$ are singlet and triplet reaction rates, respectively. $r_A$ and $r_B$ are the spin relaxation rates of radicals A and B, respectively. The allowed region is shaded with black lines.}
      \label{Fig:Triplet1}
\end{figure}

Regions in our parameter space where simulated behavior is per the experimental observation (i.e., a positive change in $\Phi^{(T)}_{T}$ at 500 $\mu T$  and a negative change at 200 $\mu T$ with respect to geomagnetic control) is simply referred to as the allowed region. In Fig.~\ref{Fig:Triplet1}, this region in $k_{S}-k_{T}$ plane is identified (shaded with black lines) for two pairs of spin relaxation rates ($r_A=10^5$ $s^{-1};r_B=10^5$ $ s^{-1}$ and $r_A=10^5$ $ s^{-1};r_B=10^6$ $ s^{-1}$). The colored shading represents the percentage changes in triplet yield at 200 $\mu T$ with respect to geomagnetic control, and the white contour lines represent the percentage changes in triplet yield at 500 $\mu T$ with respect to geomagnetic control. The allowed region in Fig.~\ref{Fig:Triplet1a} could be divided into two sub-regions: (i) vertical region centered around $k_S=3\times10^6$ $s^{-1}$ and (ii) horizontal region centered around $k_T=1\times10^6$ $s^{-1}$.  

\par
\begin{figure}[ht!]
     \centering
     \begin{subfigure}[b]{0.80\textwidth}
         \centering
         \includegraphics[width=\textwidth]{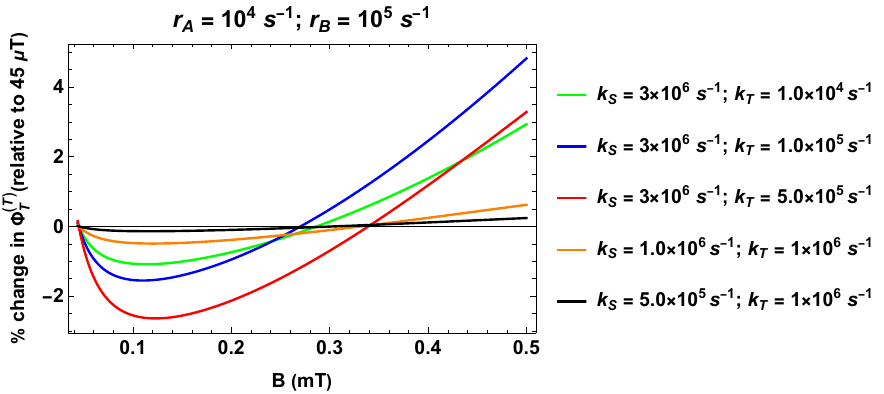}
         \caption{}
         \label{Fig:Triplet3a}
     \end{subfigure}
     \hfill
     \begin{subfigure}[b]{0.80\textwidth}
         \centering
         \includegraphics[width=\textwidth]{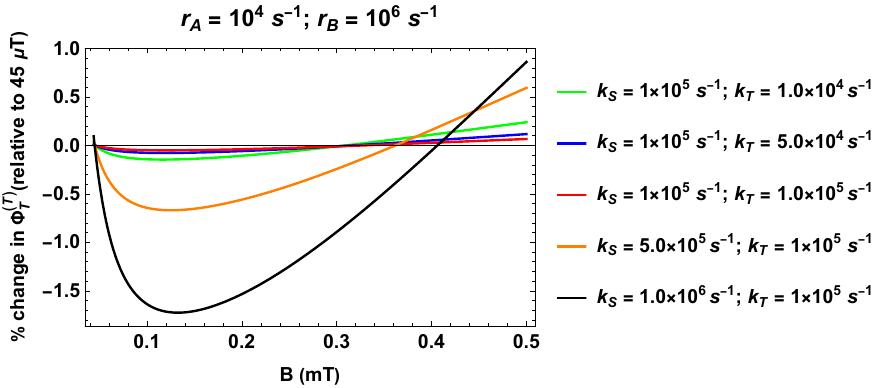}
         \caption{}
         \label{Fig:Triplet3b}
     \end{subfigure}
     
      \caption{Percentage change in the fractional triplet yield for triplet-born RP with respect to the geomagnetic control (45 $\mu T$) as a function of the magnetic field. The HFCC value $a_1$ is taken to be $802.9$ $\mu T$. (a) $r_A=10^4$ $s^{-1}$ and $r_B=10^5$ $s^{-1}$. (b) $r_A=10^4$ $s^{-1}$ and $r_B=10^6$ $s^{-1}$. $k_S$ and $k_T$ are singlet and triplet reaction rates, respectively. $r_A$ and $r_B$ are the spin relaxation rates of radicals A and B, respectively. }
        \label{Fig:Triplet3}
\end{figure}

Fig.~\ref{Fig:Triplet3} is the counter part of Fig. 3 of the manuscript for $r_A=10^4$ $s^{-1}$.

\subsection*{Singlet-born radical pair}
\begin{figure}[ht!]
     \centering

     \begin{subfigure}[b]{0.45\textwidth}
         \centering
         \includegraphics[width=\textwidth]{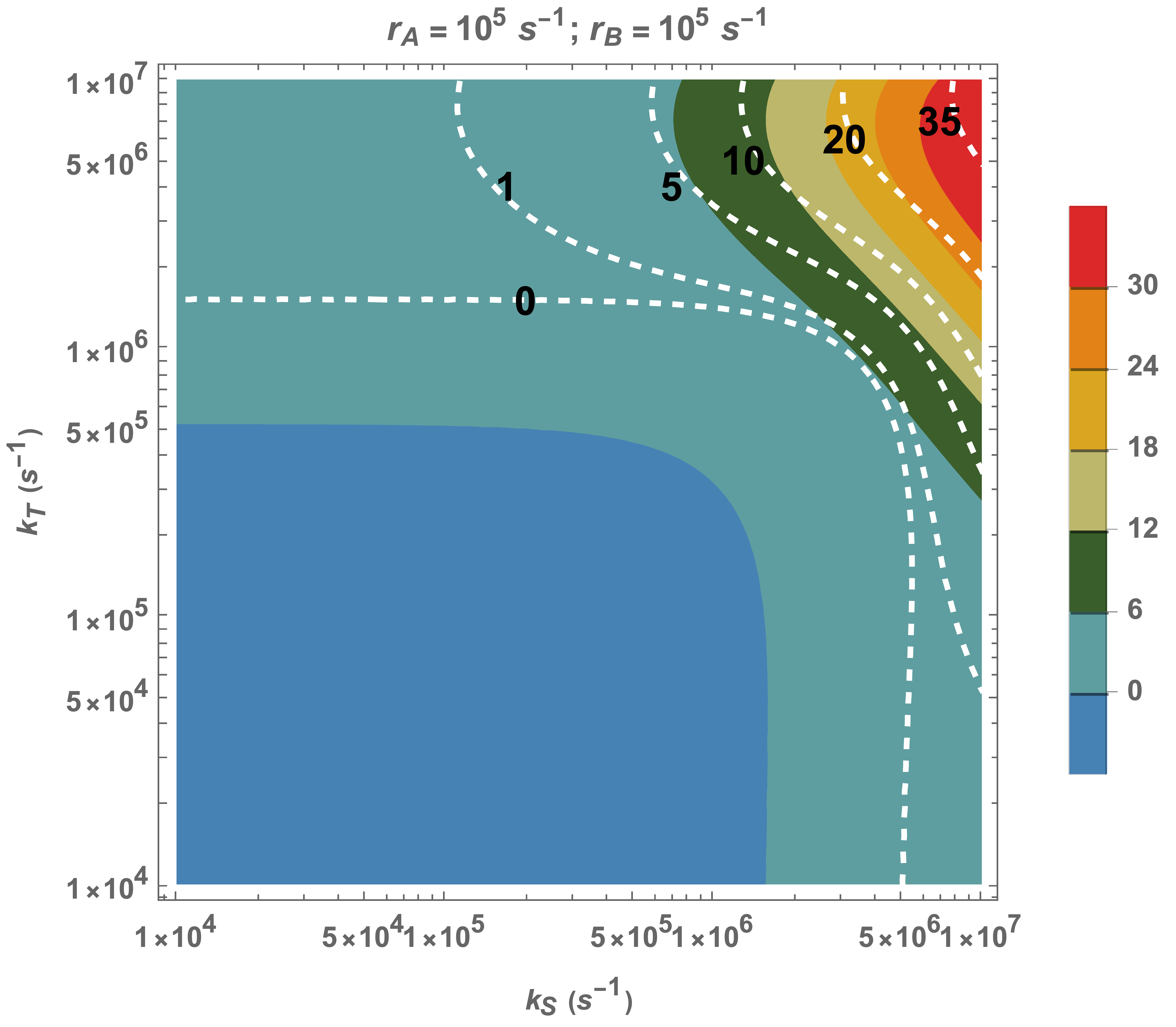}
         \caption{}
         \label{Fig:Singlet1a}
     \end{subfigure}
     \hfill
     \begin{subfigure}[b]{0.45\textwidth}
         \centering
         \includegraphics[width=\textwidth]{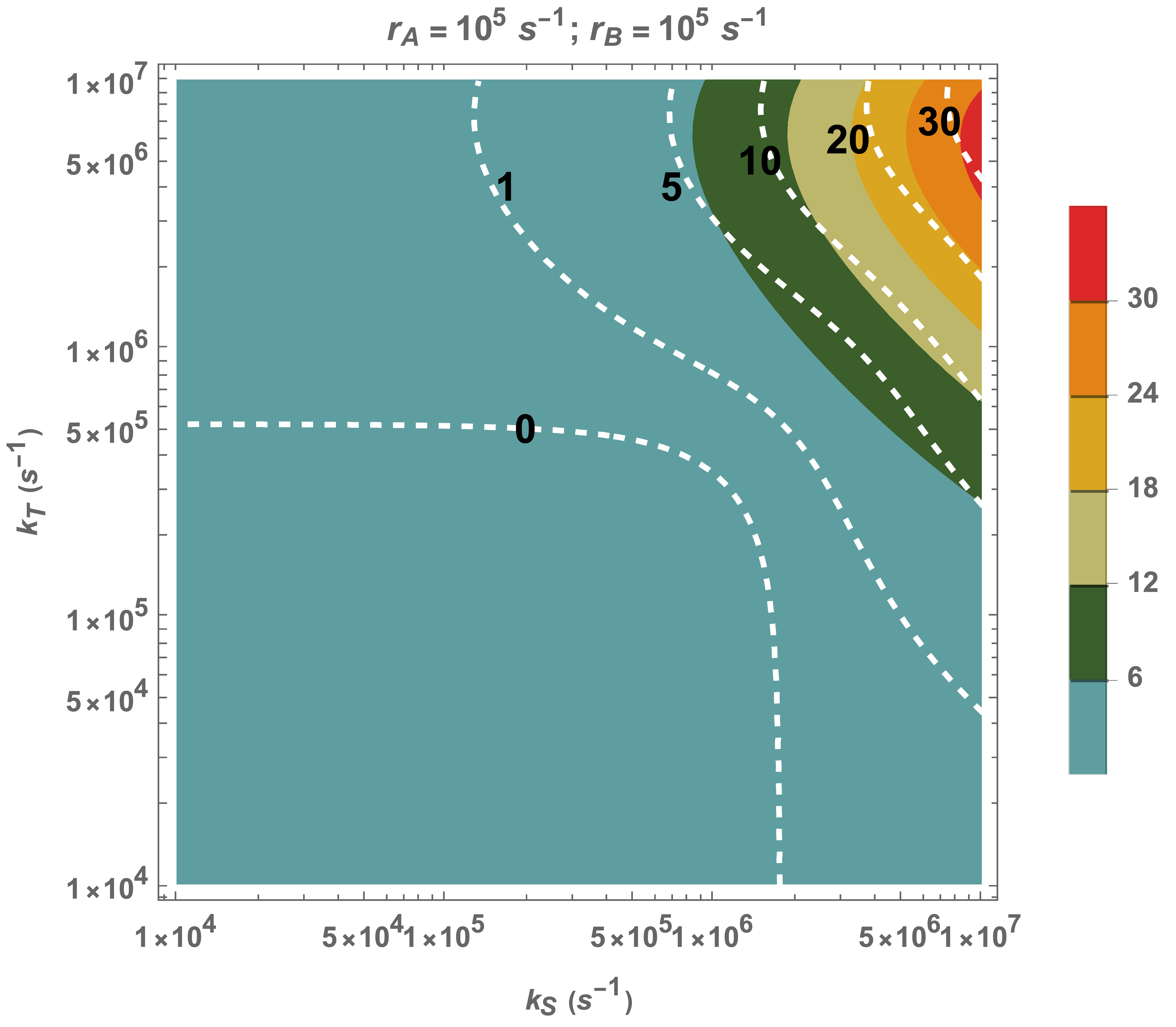}
         \caption{}
         \label{Fig:Singlet1b}
     \end{subfigure}
    
      \caption{The triplet yield changes for singlet-born RP with respect to control (45 $\mu T$) at 200 $\mu T$ and 500 $\mu T$ in $k_{S}-k_{T}$ plane. The colored shading represents the percentage changes in triplet yield at 200 $\mu T$ with respect to geomagnetic control, and the white contour lines represent the percentage changes in triplet yield at 500 $\mu T$ with respect to geomagnetic control. The HFCC value $a_1$ is taken to be $802.9$ $\mu T$. (a) $r_A=10^5$ $ s^{-1}$ and $r_B=10^5$ $  s^{-1}$. (b) $r_A=10^5$ $ s^{-1}$ and $r_B=10^6$ $ s^{-1}$. $k_S$ and $k_T$ are singlet and triplet reaction rates, respectively. $r_A$ and $r_B$ are the spin relaxation rates of radicals A and B, respectively.}
      \label{Fig:Singlet1}
\end{figure}

Fig.~\ref{Fig:Singlet1} shows the size of triplet yield changes in $k_{S}-k_{T}$ plane for singlet-born RP with respect to control (45 $\mu T$) at 200 $\mu T$ and 500 $\mu T$ for two pairs of spin relaxation rates ($r_A=10^5$ $ s^{-1};r_B=10^5$ $ s^{-1}$ and $r_A=10^5$ $ s^{-1};r_B=10^6$ $ s^{-1}$). As before, the colored shading represents the percentage changes in triplet yield at 200 $\mu T$ with respect to geomagnetic control, and the white contour lines represent the percentage changes in triplet yield at 500 $\mu T$ with respect to geomagnetic control. 

\subsection*{F-pair}
\begin{figure}[ht!]
     \centering

     \begin{subfigure}[b]{0.45\textwidth}
         \centering
         \includegraphics[width=\textwidth]{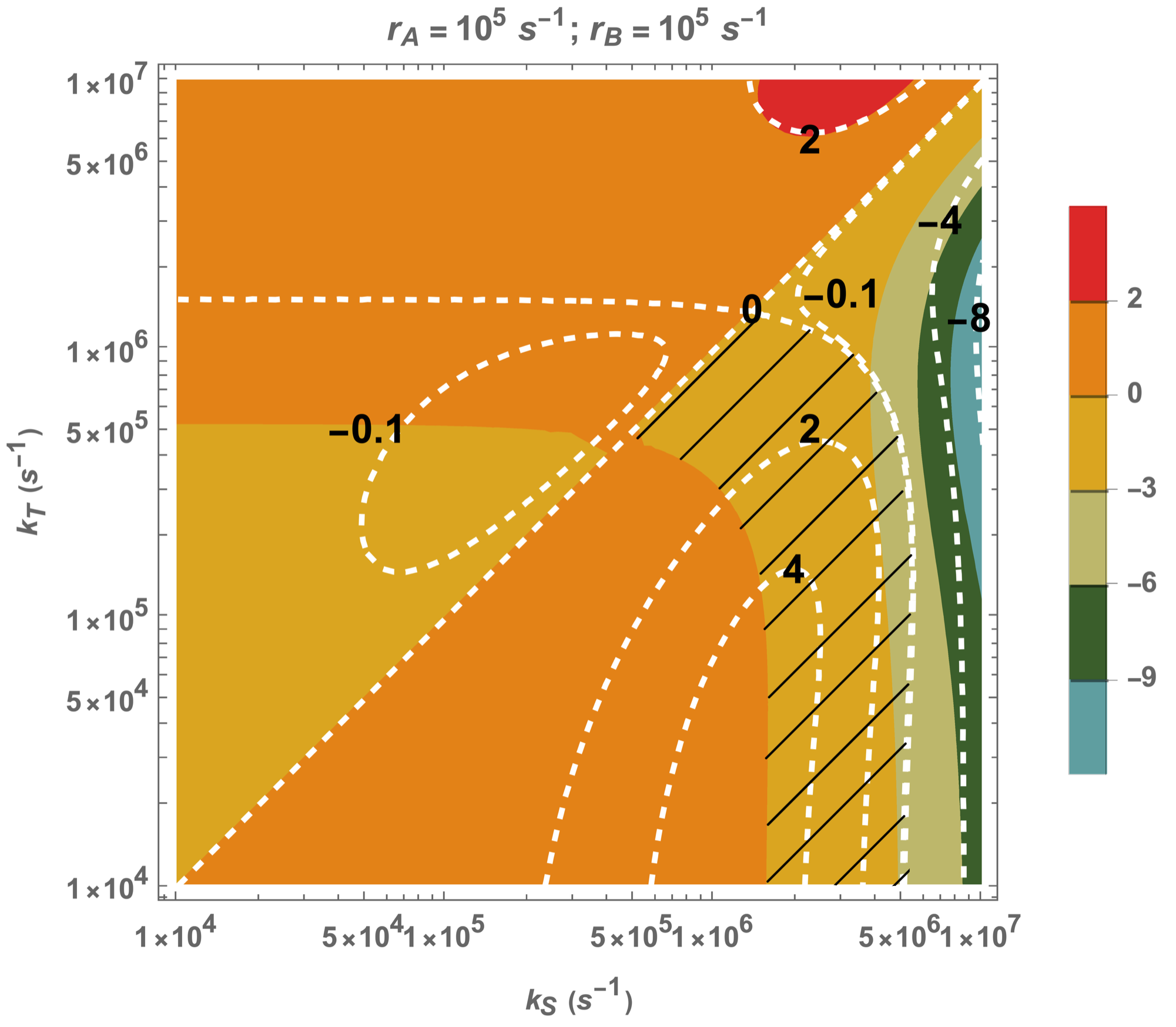}
         \caption{}
         \label{Fig:Fpair1a}
     \end{subfigure}
     \hfill
     \begin{subfigure}[b]{0.45\textwidth}
         \centering
         \includegraphics[width=\textwidth]{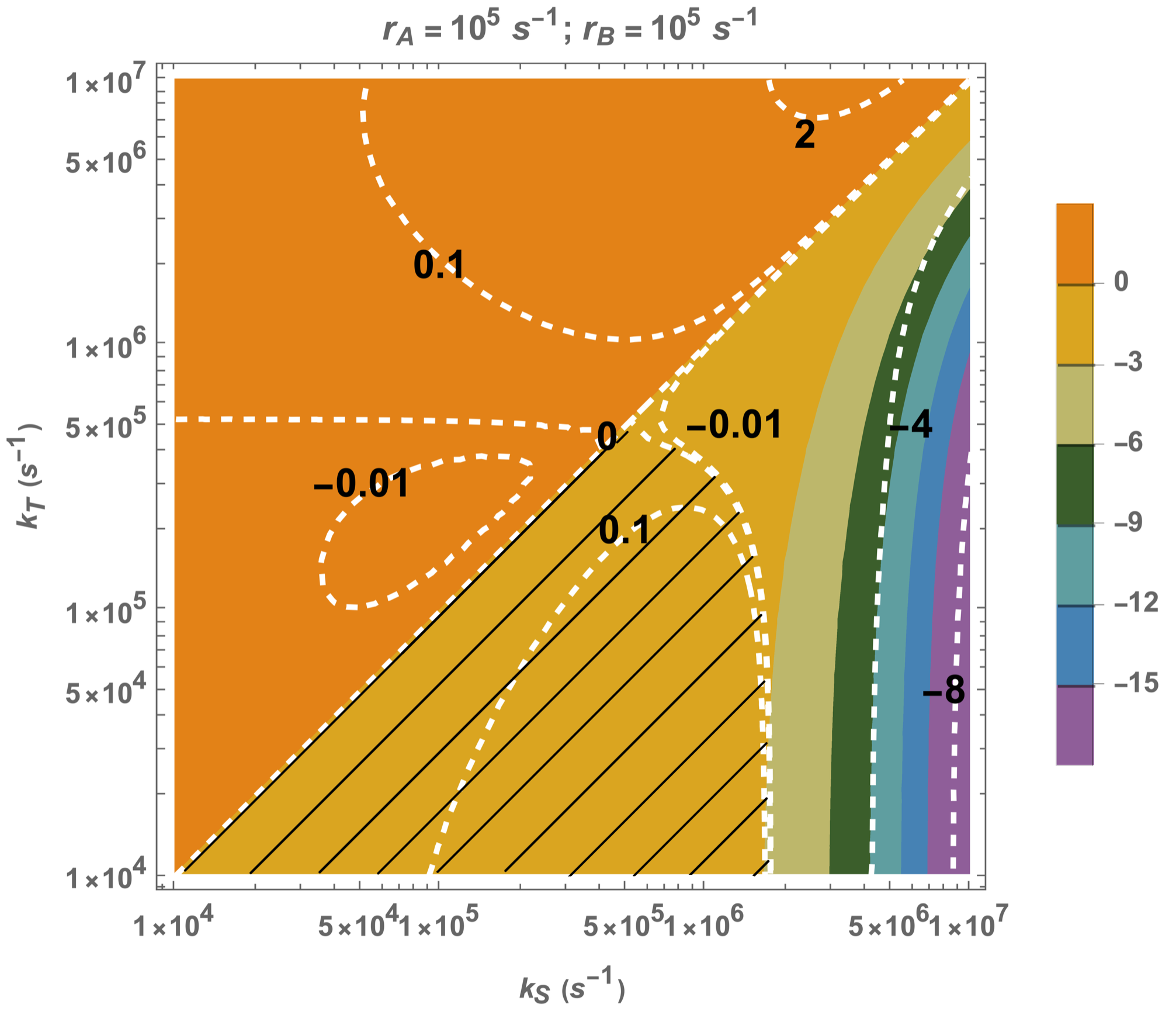}
         \caption{}
         \label{Fig:Fpair1b}
     \end{subfigure}
    
      \caption{The triplet yield changes for RPs that start as F-pairs at 200 $\mu T$ and 500 $\mu T$ with respect to control (45 $\mu T$) in $k_{S}-k_{T}$ plane. The colored shading represents the percentage changes in triplet yield at 200 $\mu T$ with respect to geomagnetic control, and the white contour lines represent the percentage changes in triplet yield at 500 $\mu T$ with respect to geomagnetic control. The HFCC value $a_1$ is taken to be $802.9$ $\mu T$. (a) $r_A=10^5$ $ s^{-1}$ and $r_B=10^5$ $ s^{-1}$. (b) $r_A=10^5$ $ s^{-1}$ and $r_B=10^6$ $ s^{-1}$. The allowed region is shaded with black lines. $k_S$ and $k_T$ are singlet and triplet reaction rates, respectively. $r_A$ and $r_B$ are the spin relaxation rates of radicals A and B, respectively.}
      \label{Fig:Fpair1}
\end{figure}

The percentage of triplet yield changes for RPs that start as F-pairs with respect to control (45 $\mu T$) at 200 $\mu T$ and 500 $\mu T$ in $k_{S}-k_{T}$ plane are shown in Fig.~\ref{Fig:Fpair1} for two pairs of spin relaxation rates ($r_A=10^5$ $ s^{-1};r_B=10^5$ $ s^{-1}$ and $r_A=10^5$ $ s^{-1};r_B=10^6$ $ s^{-1}$). 

\begin{figure}[ht!]
     \centering
         \centering
         \includegraphics[width=0.60\textwidth]{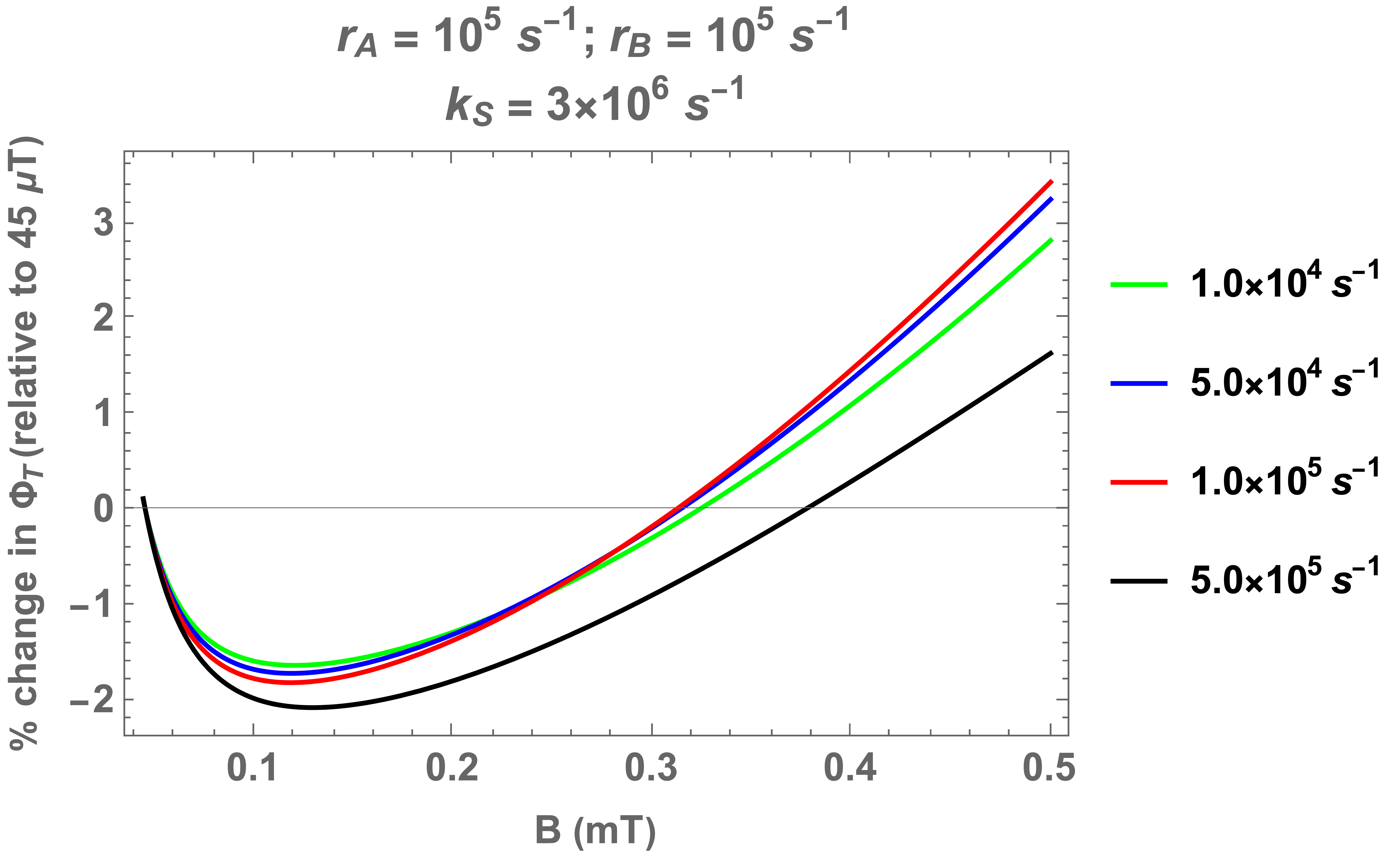}

      \caption{Percentage change in the fractional triplet yield for RP  that start as F-pair with respect to the geomagnetic control (45 $\mu T$) as a function of the magnetic field. The HFCC value $a_1$ is taken to be $802.9$ $\mu T$. $r_A=10^5$ $s^{-1}$ and $r_B=10^5$ $s^{-1}$. $k_S$ is fixed at $3\times10^6$ $s^{-1}$ and $k_T$ is varied. $k_S$ and $k_T$ are singlet and triplet reaction rates, respectively. $r_A$ and $r_B$ are the spin relaxation rates of radicals A and B, respectively.}
        \label{Fig:Fpair2}
\end{figure}

Fig.~\ref{Fig:Fpair2} shows the percentage change in triplet-yield for various points in the allowed region of Fig.~\ref{Fig:Fpair1a}.
\subsection*{Exchange interaction}
\begin{figure}[ht!]
     \centering

         \includegraphics[width=0.45\textwidth]{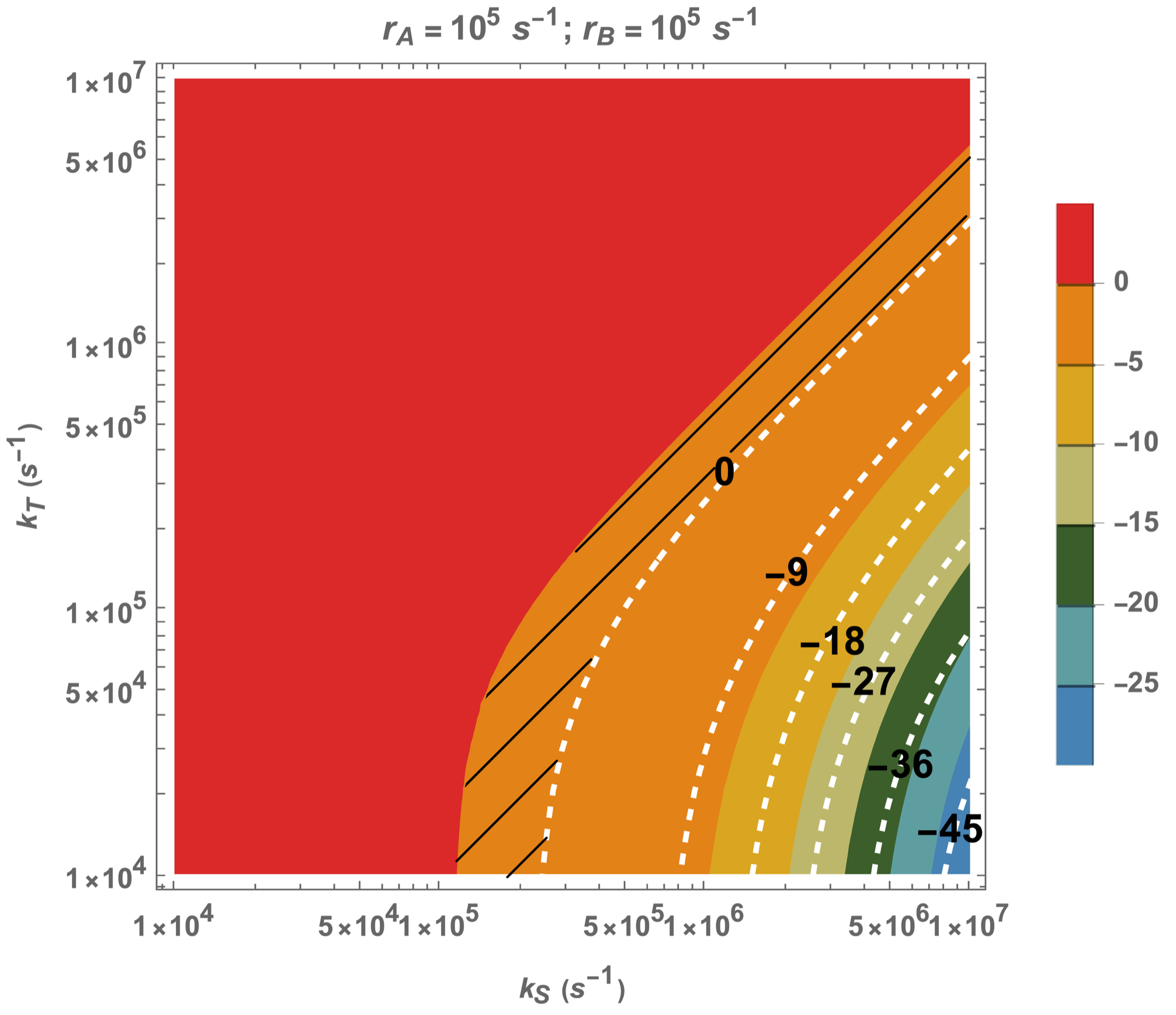}
         \caption{Exchange interaction ($J=-500$ $\mu T$). $r_A=10^5$ $ s^{-1}$ and $r_B=10^5$ $ s^{-1}$. The HFCC value $a_1$ is taken to be $802.9$ $\mu T$. The percentage of triplet-born triplet yield changes for RPs at 200 $\mu T$ and 500 $\mu T$ with respect to control (45 $\mu T$) in $k_{S}-k_{T}$ plane. The colored shading represents the percentage changes in triplet yield at 200 $\mu T$ with respect to geomagnetic control, and the white contour lines represent the percentage changes in triplet yield at 500 $\mu T$ with respect to geomagnetic control. The allowed region is shaded with black lines. $k_S$ and $k_T$ are singlet and triplet reaction rates, respectively. $r_A$ and $r_B$ are the spin relaxation rates of radicals A and B, respectively.}
         \label{Fig:Exchange1}
    \end{figure}

\begin{figure}[ht!]
     \centering
    
         \includegraphics[width=0.60\textwidth]{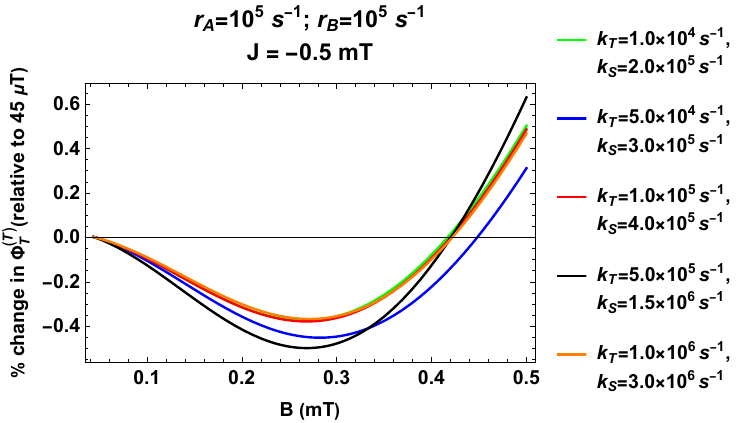}
    
      \caption{Exchange interaction ($J=-500$ $\mu T$). $r_A=10^5$ $ s^{-1}$ and $r_B=10^5$ $ s^{-1}$. The HFCC value $a_1$ is taken to be $802.9$ $\mu T$. Percentage change in the fractional triplet yield for triplet-born RP with respect to the geomagnetic control (45 $\mu T$) as a function of the magnetic field. $k_S$ and $k_T$ are singlet and triplet reaction rates, respectively. $r_A$ and $r_B$ are the spin relaxation rates of radicals A and B, respectively.}
      \label{Fig:Exchange2}
\end{figure}

 We now include an exchange interaction in our model to study the impact of non-zero inter-radical interactions. This can be done by putting $\hat{H}_{IR} = J\hat{S}_A.\hat{S}_B$, where $J$ represents the strength of the interaction. Introducing a small exchange interaction ($J=-10$ $\mu T$) does not have any significant effects. Intermediate interaction strengths, on the other hand, have drastic effects. We could not find any region in the $k_S$-$k_T$ plane for which $\Phi^{(T)}_{S}$ changes in accordance with experimental observations for values between $J=-15$ $\mu T$ and $J=-220$ $\mu T$. However, the allowed region starts reappearing around $J=-220$ $\mu T$. We now discuss the case for $J=-500$ $\mu T$. The size of triplet yield changes in $k_{S}-k_{T}$ plane for singlet-born RP with respect to control (45 $\mu T$) at 200 $\mu T$ and 500 $\mu T$ are plotted in Fig.~\ref{Fig:Exchange1}. In Fig.~\ref{Fig:Exchange2}, we have also plotted the percentage change in the $\Phi^{(T)}_{T}$ with respect to the control scenario as a function of the magnetic field for a few representative parameter values from this region. As can be seen, the effects of about half a percent can be reached. These results are for $r_A=10^5$ and $r_B=10^5$. Similar results are observed if $r_B$ is increased to $10^6$.

\subsection*{Multiple hyperfine interactions}
\begin{figure}[ht!]
     \centering

         \includegraphics[width=0.45\textwidth]{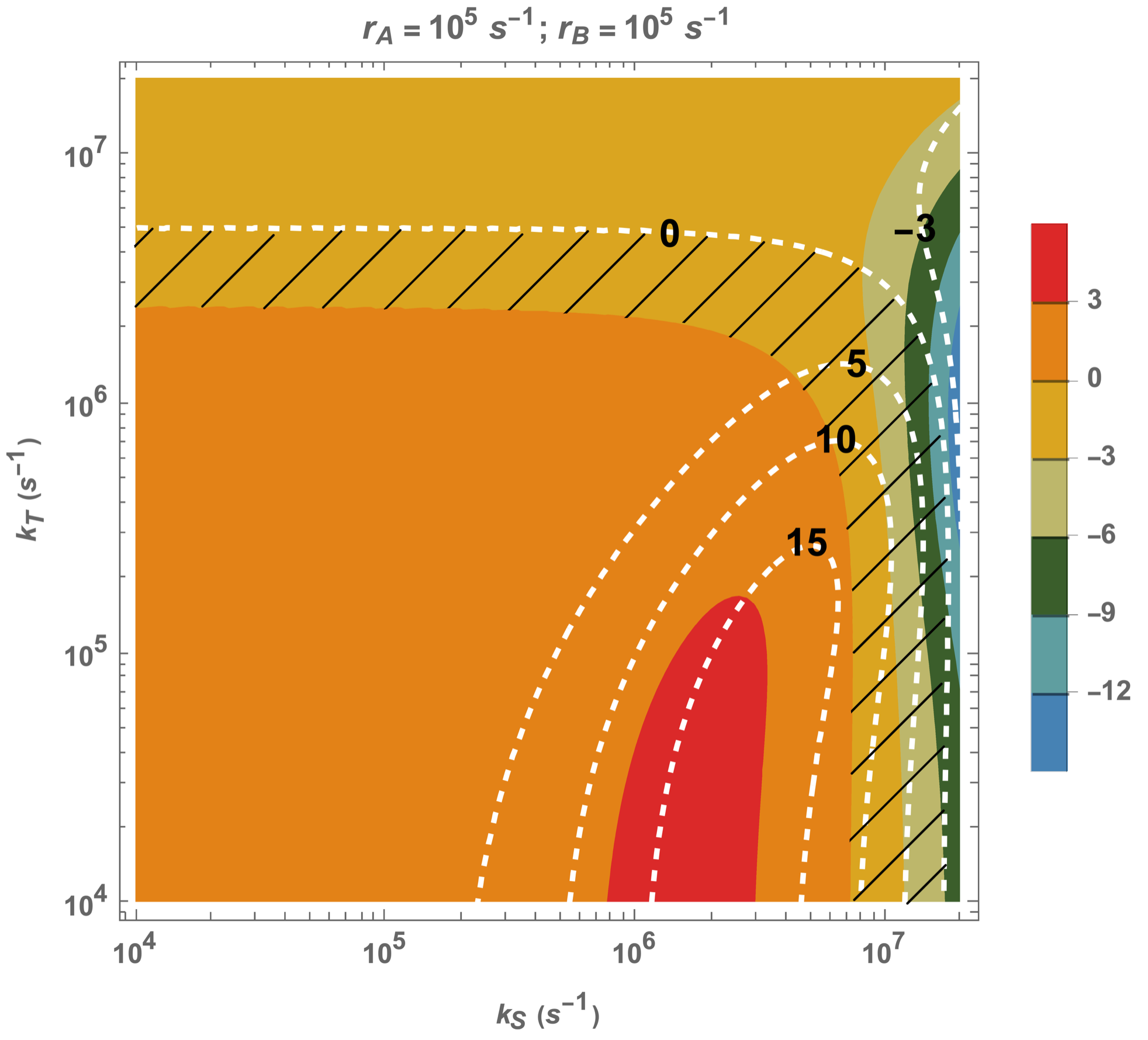}

      \caption{Multiple HF interactions (two: H5 and N5). $r_A=10^5$ $s^{-1}$ and $r_B=10^5$ $s^{-1}$. $a_1=802.9$ $\mu T$ and $a_2=431.3$ $\mu T$ . The percentage of triplet-born triplet yield changes for RPs at 200 $\mu T$ and 500 $\mu T$ with respect to control (45 $\mu T$) in $k_{S}-k_{T}$ plane. The colored shading represents the percentage changes in triplet yield at 200 $\mu T$ with respect to geomagnetic control, and the white contour lines represent the percentage changes in triplet yield at 500 $\mu T$ with respect to geomagnetic control. The allowed region is shaded with black lines. $k_S$ and $k_T$ are singlet and triplet reaction rates, respectively. $r_A$ and $r_B$ are the spin relaxation rates of radicals A and B, respectively.}
       \label{Fig:HFI1}
\end{figure}

\begin{figure}[ht!]
     \centering
         \centering
         \includegraphics[width=0.6\textwidth]{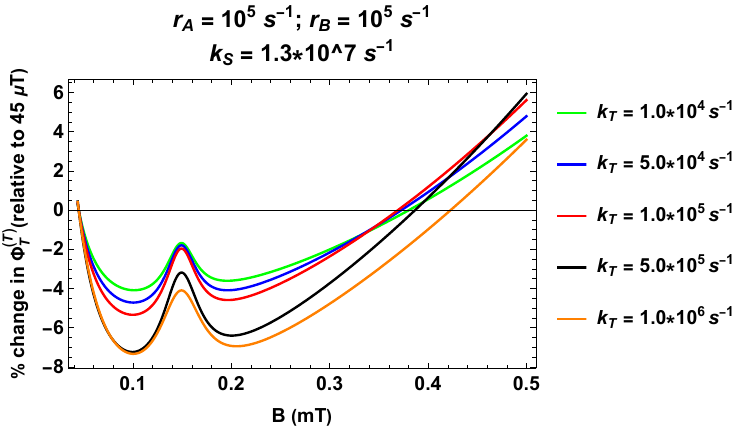}

      \caption{ Multiple HF interactions (two: H5 and N5). $a_1=802.9$ $\mu T$ and $a_2=431.3$ $\mu T$ .Percentage change in the fractional triplet yield for triplet-born RP with respect to the geomagnetic control (45 $\mu T$) as a function of the magnetic field. $r_A=10^5$ $ s^{-1}$ and $r_B=10^5$ $ s^{-1}$. $k_S$ is fixed at $1.3\times10^7$  $s^{-1}$ and $k_T$ is varied. $k_S$ and $k_T$ are singlet and triplet reaction rates, respectively. $r_A$ and $r_B$ are the spin relaxation rates of radicals A and B, respectively.}
        \label{Fig:HFI2}
\end{figure}

We have considered only one HF interaction (with H5) for \ch{FH^{.}}. In reality, the situation is more complex, and this radical interacts with multiple nuclear spins. These additional interactions can affect the spin dynamics of the RP. To study the impact of having multiple HF interactions, we plotted the counterpart of Fig.~\ref{Fig:Triplet1} but with one additional nuclear spin (N5, with HFCC $a_2=431.3$ $\mu T$) (see Fig.~\ref{Fig:HFI1}). The main effect of this change is to shift the allowed region to slightly higher rate values. Fig.~\ref{Fig:HFI2} shows the percentage change in $\Phi^{(T)}_{T}$ with respect to the geomagnetic control as a function of the magnetic field for parameter values in the vertical allowed region of Fig.~\ref{Fig:HFI1}. $k_S$ value is fixed at $1.3\times10^7$ $ s^{-1}$ and $k_T$ is varied. Comparison with Fig. 3a of the manuscript shows that incorporating this second HF interaction introduces some wiggles at 150 $\mu T$. The behavior is qualitatively similar to the case of one HF interaction at 200 and 500 $\mu T$. The effect size is in the same order as in Fig.3a of the manuscript. All these figures are plotted for $r_A=10^5$ $ s^{-1}$ and $r_B=10^5$ $ s^{-1}$. Similar results can be obtained if $r_B$ is set at $10^6$ $s^{-1}$. For more nuclei, the wiggles can be larger for smaller $k_T$ values. From these observations, it may be concluded that having more than one HF interaction does not change our main conclusions.


\subsection*{Bound vs free flavin}
Ramsay and Kattnig~\cite{ramsay2022radical} have suggested that in the case of bound flavin, the correct approach would be to take an anisotropic HF coupling tensor. Then, the yields should be averaged over the orientation of the magnetic field with respect to the radicals. To this aim, we calculated triplet yield by averaging over 100 uniformly distributed magnetic field orientations. The result of the simulation shows that for triplet-born RPs, there is always a positive change for both 500 $\mu T$  and 200 $\mu T$ relative to the control. For singlet-born RP, the change is always negative. Hence, these results indicate that flavin in question might not be protein-bound but rather free.

\subsection*{Isotopic effects of oxygen}
Isotope effects can indicate the RPM\cite{smith2021radical,zadeh2021Li}. Naturally occurring oxygen is almost exclusively found in the form of the \ch{^{16}O} isotope, which has a zero spin. If one of the oxygen atoms in superoxide radical is replaced with \ch{^{17}O}, which has a spin, $I_{3} = 5/2$, an additional HF term must be added to the RP Hamiltonian. The new HF part of Hamiltonian for our RP system reads as follows:
\begin{equation}\label{eq:ham2}
\centering
\hat{H_{HFI}} =  a_{1}\mathbf{\hat{S}}_{A}.\mathbf{\hat{I}_{1}} + a_{3}\mathbf{\hat{S}}_{B}.\mathbf{\hat{I}_{3}}, 	
\end{equation}
$\mathbf{\hat{I}_{3}}$ is the nuclear spin operator of the \ch{^{17}O}, $a_{3}$ is the HFCC between the \ch{^{17}O} and the radical electron B. We take the value of $a_{3}$ to be $1886.8$ $\mu T$~\cite{rishabh2022radical}. 
\par
\begin{figure}[ht!]
     \centering

         \includegraphics[width=0.6\textwidth]{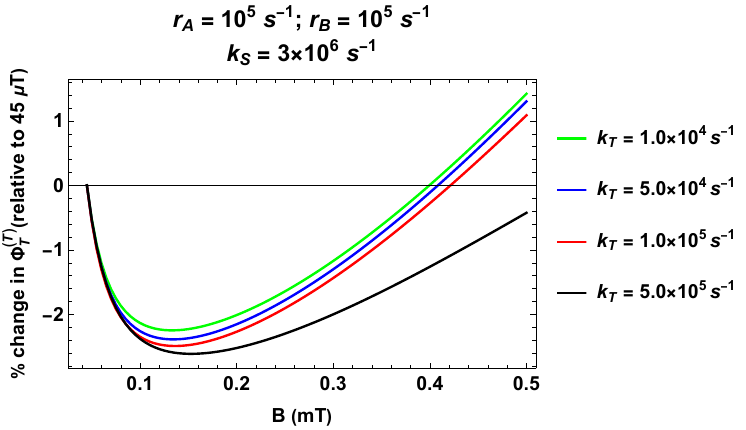}

      \caption{Substitution with \ch{^{17}O}. $a_1=802.9$ $\mu T$ and $a_3=1886.8$ $\mu T$ .Percentage change in the fractional triplet yield for triplet-born RP with respect to the geomagnetic control (45 $\mu T$) as a function of the magnetic field. $r_A=10^5$ $ s^{-1}$ and $r_B=10^5$ $ s^{-1}$. $k_S$ is fixed at $3\times10^6$ $s^{-1}$ and $k_T$ is varied. $k_S$ and $k_T$ are singlet and triplet reaction rates, respectively. $r_A$ and $r_B$ are the spin relaxation rates of radicals A and B, respectively.}
       \label{Fig:O171}

\end{figure}

Considerable overlap was found between the allowed regions with and without \ch{^{17}O}, suggesting that the qualitative behavior of the WMF effects may not be affected by the introduction of \ch{^{17}O} isotope. To understand the quantitative effects of this isotopic substitution, we have plotted the percentage change in $\Phi^{(T)}_{T}$ as a function of the magnetic field for RP containing one \ch{^{17}O} for exactly the same rate values as in Fig. 3a of the manuscript (See Fig~\ref{Fig:O171}). The comparison with Fig. 3a of the manuscript reveals no significant deviation in the strength of effects at 200 $\mu T$, the effects at 500 $\mu T$ are slightly reduced, and even flips sign for $k_T = 5\times10^5$ $ s^{-1}$.

\subsection*{Radical triad model}
\begin{figure}[ht!]
     
         \centering
         \includegraphics[width=0.45\textwidth]{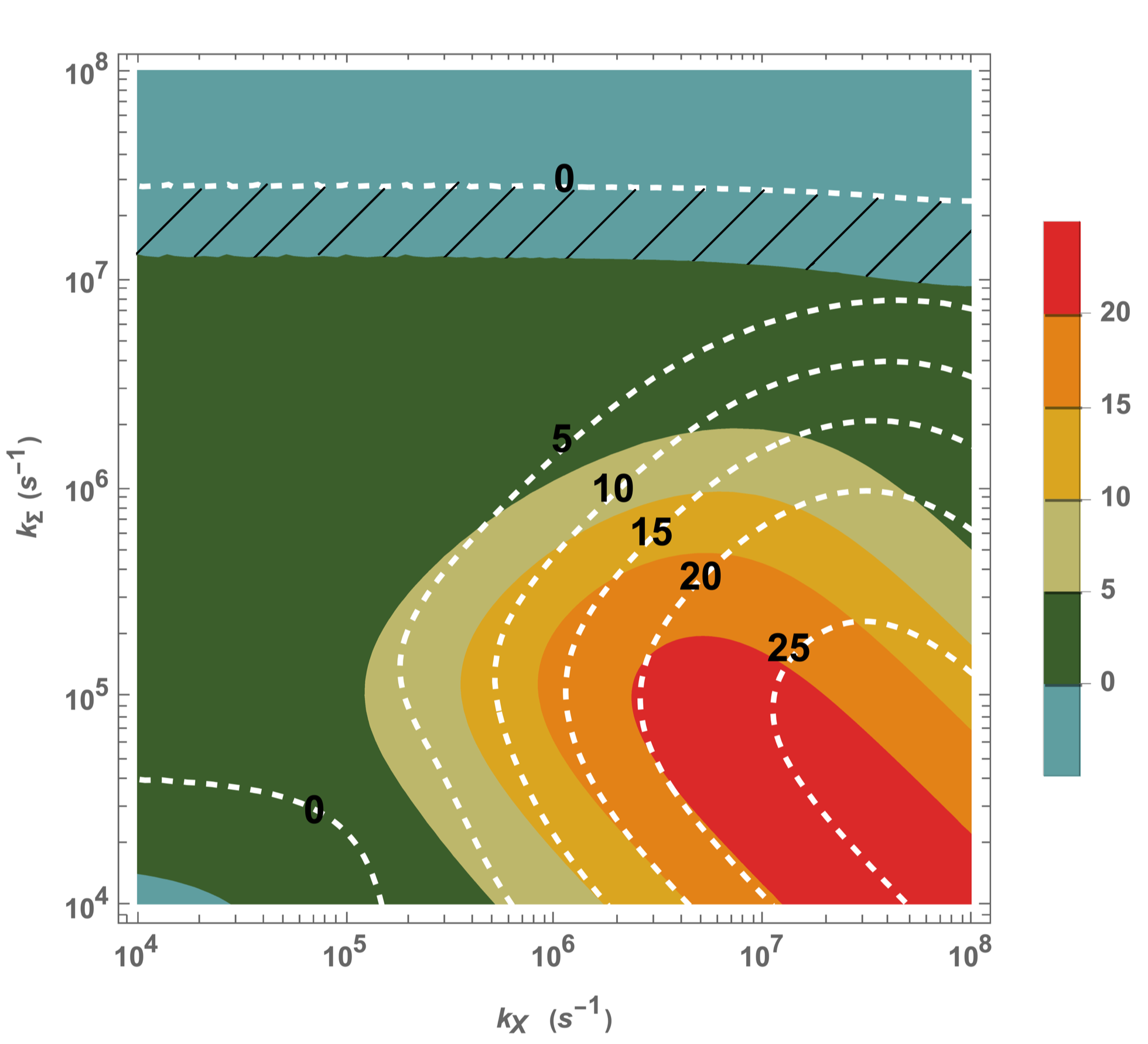}
         \caption{Radical triad model. The precentage of triplet yield changes for RPs at 200 $\mu T$ and 500 $\mu T$ with respect to control (45 $\mu T$) in $k_{X}-k_{\Sigma}$ plane. The colored shading represents the percentage changes in triplet yield at 200 $\mu T$ with respect to geomagnetic control, and the white contour lines represent the percentage changes in triplet yield at 500 $\mu T$ with respect to geomagnetic control. The HFCC value $a_1$ is taken to be $802.9$ $\mu T$, and the HFCC value for \ch{A^{.}} is taken to be $2000$ $\mu T$. The allowed region is shaded with black lines. $k_X$ and $k_\Sigma$ are defined in the above paragraph.}
         \label{Fig:Triad1}
  
\end{figure}

Kattnig's group has proposed a radical triad model involving a third scavenger radical to deal with the fast spin relaxation of superoxide radicals. Here, we present an analysis based on a simple radical triad model (see Ref.~\cite{ramsay2022radical} for details of this model). The three radicals involved are namely \ch{FH^{.}}, \ch{O2^{.-}}, and an unknown scavenger radical (\ch{A^{.}}). We assume instantaneous relaxation of superoxide, and no relaxation is assumed for the other two radicals. Further, an effective oxidation process is also assumed. Only one isotropic HF interaction is considered for \ch{FH^.} and \ch{A^{.}}. Exchange and dipolar interactions are not taken into account. There are two free parameters in our model, namely, reaction rates $k_X$ and $k_\Sigma$. $k_X$ accounts for \ch{FH^.}-\ch{A^{.}} recombination and $k_\Sigma$ is the effective depopulation rate constant of the triad system (see Ref.~\cite{ramsay2022radical}). For a suitable HF coupling constant for \ch{A^{.}}, we can find a region in the parameter space within a reasonable range of parameters for which \ch{O2^{.-}} yield changes in accordance with experimental observations (see Fig.~\ref{Fig:Triad1}).

\begin{figure}[ht!]
     
         \centering
         \includegraphics[width=0.6\textwidth]{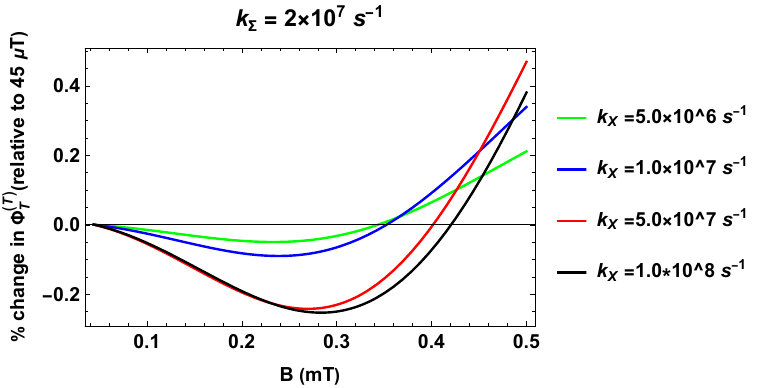}

      \caption{Radical triad model. Percentage change in the fractional triplet yield with respect to the geomagnetic control (45 $\mu T$) as a function of the magnetic field. $k_\Sigma$ is fixed at $2.0\times10^7$ $s^{-1}$, and $k_X$ is varied. $k_X$ and $k_\Sigma$ are defined in the above paragraph.}
      \label{Fig:Triad2}
\end{figure}

 Fig.~\ref{Fig:Triad2} presented here shows the percentage change in the fractional superoxide yield with respect to the control scenario (45 $\mu T$ magnetic field) as a function of the magnetic field for a representative set of parameters $k_X$ and $k_\Sigma$ in the allowed region of Fig.~\ref{Fig:Triad2}. 

\newpage


\end{document}